\documentclass{aastex61}

\newcommand\aastex{AAS\TeX}

\shorttitle{\aastex\ Unveiling the dynamical state of massive clusters through the ICL fraction}
\shortauthors{Jim\'enez-Teja et al.}

\begin{document}

\title{Unveiling the dynamical state of massive clusters through the ICL fraction}

\correspondingauthor{Yolanda Jim\'enez-Teja}
\email{yojite@iaa.es, yolanda@on.br}

\author{Yolanda Jim\'enez-Teja}
\affil{Observat\'orio Nacional, Rua General Jos\'e Cristino, 77 - Bairro Imperial de S\~ao Crist\'ov\~ao, Rio de Janeiro, 20921-400, Brazil}

\author{Renato Dupke}
\affiliation{Observat\'orio Nacional, Rua General Jos\'e Cristino, 77 - Bairro Imperial de S\~ao Crist\'ov\~ao, Rio de Janeiro, 20921-400, Brazil}
\affiliation{Department of Physics and Astronomy, University of Alabama, Box 870324, Tuscaloosa, AL 35487} 
\affiliation{Department of Astronomy, University of Michigan, 311 West Hall, 1085 South University Ave., Ann Arbor, MI 48109-1107}
\affiliation{Eureka Scientific Inc., 2452 Delmer St, Suite 100, Oakland, CA 94602} 

\author{Narciso Ben\'itez}
\affiliation{Instituto de Astrof\'isica de Andaluc\'ia (CSIC), Glorieta de la Astronom\'ia s/n, Granada, 18008, Spain}

\author{Anton M. Koekemoer}
\affiliation{Space Telescope Science Institute, 3700 San Martin Dr., Baltimore, MD, 21218, USA}

\author{Adi Zitrin}
\affiliation{Physics Department, Ben-Gurion University of the Negev, P.O. Box 653, Be'er-Sheva 8410501, Israel}

\author{Keiichi Umetsu}
\affiliation{Institute of Astronomy and Astrophysics, Academia Sinica, P.O. Box 23-141, Taipei 10617, Taiwan}

\author{Bodo L. Ziegler}
\affiliation{Department of Astrophysics, University of Vienna, T\"urkenschanzstrasse 17, 1180 Vienna, Austria}

\author{Brenda L. Frye}
\affiliation{Department of Astronomy, Steward Observatory, University of Arizona, 933 North Cherry Avenue, Tucson, AZ, 85721, USA}

\author{Holland Ford}
\affiliation{Department of Physics and Astronomy, The Johns Hopkins University Homewood Campus, Baltimore, MD 21218, USA}

\author{Rychard J. Bouwens}
\affiliation{Leiden Observatory, Leiden University, NL-2300 RA Leiden, The Netherlands}

\author{Larry D. Bradley}
\affiliation{Space Telescope Science Institute, 3700 San Martin Dr., Baltimore, MD, 21218, USA}

\author{Thomas Broadhurst}
\affiliation{Fisika Teorikoa, Zientzia eta Teknologia Fakultatea, Euskal Herriko Unibertsitatea UPV/EHU, E-48080 Bilbao, Spain}
\affiliation{IKERBASQUE, Basque Foundation for Science, Alameda Urquijo, 36-5, E-48008 Bilbao, Spain}

\author{Dan Coe}
\affiliation{Space Telescope Science Institute, 3700 San Martin Dr., Baltimore, MD, 21218, USA}

\author{Megan Donahue}
\affiliation{Physics and Astronomy Department, Michigan State University, East Lansing, MI, 48824 USA}

\author{Genevieve J.  Graves}
\affiliation{Department of Astronomy, University of California, Berkeley, CA 94720, USA}

\author{Claudio Grillo}
\affiliation{Dark Cosmology Centre, Niels Bohr Institute, University of Copenhagen, Juliane Maries Vej 30, DK-2100 Copenhagen, Denmark}

\author{Leopoldo Infante}
\affiliation{Instituto de Astrof\'isica y Centro de Astroingenier\'ia, Facultad de F\'isica, Pontificia Universidad Cat\'olica de Chile, Vicu\~na Mackenna 4860, 7820436 Macul, Santiago, Chile}

\author{Stephanie Jouvel}
\affiliation{Department of Physics and Astronomy, University College London, 132 Hampstead road, London NW1 2PS, UK}

\author{Daniel D. Kelson}
\affiliation{The Observatories of the Carnegie Institution for Science, 813 Santa Barbara St., Pasadena, CA 91101, USA}

\author{Ofer Lahav}
\affiliation{Department of Physics and Astronomy, University College of London, Gower Street, London WC1E 6BT, UK}

\author{Ruth Lazkoz}
\affiliation{Fisika Teorikoa, Zientzia eta Teknologia Fakultatea, Euskal Herriko Unibertsitatea UPV/EHU, E-48080 Bilbao, Spain}

\author{Dorom Lemze}
\affiliation{Department of Physics and Astronomy, The Johns Hopkins University Homewood Campus, Baltimore, MD 21218, USA}

\author{Dan Maoz}
\affiliation{School of Physics and Astronomy, Tel-Aviv University, Tel-Aviv 6997801, Israel}

\author{Elinor Medezinski}
\affiliation{Department of Astrophysical Sciences, Princeton University, Princeton, NJ 08544, USA}

\author{Peter Melchior}
\affiliation{Department of Astrophysical Sciences, Princeton University, Princeton, NJ 08544, USA}

\author{Massimo Meneghetti}
\affiliation{Osservatorio Astronomico di Bologna, Via Gobetti 93/3, 40129, Bologna, Italy}
\affiliation{INFN, Sezione di Bologna, viale Berti Pichat 6/2, I-40127 Bologna, Italy}

\author{Amata Mercurio}
\affiliation{INAF-Osservatorio Astronomico di Capodimonte, Salita Moiariello 16 I-80131 Napoli, Italy}

\author{Julian Merten}
\affiliation{Department of Physics, University of Oxford, Keble Road, Oxford OX1 3RH, U.K.}

\author{Alberto Molino}
\affiliation{Instituto de Astronomia, Geof\'isica e Ci\^encias Atmosf\'ericas, Universidade de S\~ao Paulo, 05508-090 S\~ao Paulo, Brazil}
\affiliation{Instituto de Astrof\'isica de Andaluc\'ia (CSIC), Glorieta de la Astronom\'ia s/n, Granada, 18008, Spain}

\author{Leonidas A. Moustakas}
\affiliation{Jet Propulsion Laboratory, California Institute of Technology, MS 169-506, 4800 Oak Grove Drive, Pasadena, CA 91109, USA}

\author{Mario Nonino}
\affiliation{INAF-Osservatorio Astronomico di Trieste, via G. B. Tiepolo 11, 34143 Trieste, Italy}

\author{Sara Ogaz}
\affiliation{Department of Physics and Astronomy, The Johns Hopkins University Homewood Campus, Baltimore, MD 21218, USA}

\author{Adam G. Riess}
\affiliation{Space Telescope Science Institute, 3700 San Martin Dr., Baltimore, MD, 21218, USA}
\affiliation{Department of Physics and Astronomy, The Johns Hopkins University Homewood Campus, Baltimore, MD 21218, USA}

\author{Piero Rosati}
\affiliation{Dipartimento di Fisica e Scienze della Terra, Universit\`a degli Studi di Ferrara, Via Saragat 1, I-44122 Ferrara, Italy}

\author{Jack Sayers}
\affiliation{Division of Physics, Math, and Astronomy, California Institute of Technology, Pasadena, CA 91125, USA}

\author{Stella Seitz}
\affiliation{University Observatory Munich, Scheinerstrasse 1, D-81679 Munich, Germany}
\affiliation{Max Planck Institute for Extraterrestrial Physics, Giessenbachstrasse, D-85748 Garching, Germany}

\author{Wei Zheng}
\affiliation{Department of Physics and Astronomy, The Johns Hopkins University Homewood Campus, Baltimore, MD 21218, USA}



\begin{abstract}

We have selected a sample of eleven massive clusters of galaxies observed by the Hubble Space Telescope in order to study the impact of the dynamical state on the IntraCluster Light (ICL) fraction, the ratio of total integrated ICL to the total galaxy member light. With the exception of the Bullet cluster, the sample is drawn from the Cluster Lensing and Supernova Survey and the Frontier Fields program, containing five relaxed and six merging clusters. The ICL fraction is calculated in three optical filters using the CHEFs IntraCluster Light Estimator, a robust and accurate algorithm free of a priori assumptions. We find that the ICL fraction in the three bands is, on average, higher for the merging clusters, ranging between $\sim7-23\%$, compared with the $\sim 2-11\%$ found for the relaxed systems. We observe a nearly constant value (within the error bars) in the ICL fraction of the regular clusters at the three wavelengths considered, which would indicate that the colors of the ICL and the cluster galaxies are, on average, coincident and, thus, their stellar populations. However, we find a higher ICL fraction in the F606W filter for the merging clusters, consistent with an excess of lower-metallicity/younger stars in the ICL, which could have migrated violently from the outskirts of the infalling galaxies during the merger event.

\end{abstract}




\section{Introduction} \label{sec_intro}

The Intracluster Light (ICL) is starting to get some attention for its ability to give insights into the processes driving galaxy cluster evolution. Defined as the light of the stars that do not belong to any of the galaxies of the clusters but are gravitationally bound to the potential of the system, the ICL origin, general properties and evolution are almost completely unknown. Its contribution to the total luminosity of the cluster can be significant, ranging from 10 to 50\% of the total light, where the upper limit was claimed by \cite{bernstein} in the core of the Coma cluster. This contribution is known as the ICL luminosity fraction (ICL fraction, hereafter), and its formally defined as the ratio between the ICL and the total luminosity of the cluster. The total luminosity comprises the ICL and the light from the galaxy members of the cluster. Although this parameter is conceptually very simple, accurate measurements of  the ICL are not trivial to obtain, which largely explains the scatter in the results reported in  different studies in the literature with inconsistent methodologies. Disentangling the ICL from the light of the stars in the galaxies is not straightforward, being especially complex in the case of the brightest cluster galaxy (BCG). Traditional methods are often ambiguous, relying on different a priori hypothesis that tie conditions to the final measurements, leading to different conclusions on the nature of the ICL and its properties. Assuming that the ICL formation is mainly driven by ongoing processes (e.g. tidal stripping or total disruption of dwarf galaxies) it is predicted that the ICL  fraction  will grow with decreasing redshift \citep{willman2004,krick07,burke2012,rudick2006,rudick2011}. However, several works failed to find any significant redshift dependence of the ICL or the ICL fraction, probed at different redshift ranges \citep{krick07,morishita2017,montes2017, guennou2012}.  In addition, it has been suggested that the ICL fraction should be related to the dynamical evolutionary stage of the cluster, such as a merger, since it is expected that the amount of ICL would increase with the infall material on the cluster \citep{pierini2008, adami2004,adami2013}. In fact, although involving a smaller scale, some authors already reported a correlation between the fraction of diffuse light and the dynamical state in groups, finding that the IntraGroup Light (IGL) fraction was higher in the case  of active groups (e.g., \cite{darocha2005,darocha2008}).\\

The major difficulty in looking for quantitative and qualitative relations between the ICL and other cluster properties comes from a combination of small number statistics of clusters with measured ICL, the different data quality and the use of very disparate methodology. In particular, the latter makes direct comparisons  of results from different works extremely difficult, since we cannot determine if the conclusions are in fact real or the result of a bias induced by the different techniques used, the different hypothesis assumed, or selection effects. The aim of this work is to study in a consistent way the role of the cluster's dynamics in the ICL formation. We selected a significant sample of massive merging and relaxed clusters spanning a redshift range of $0.18<z<0.54$, which is of special interest since several works suggest that the ICL is mainly formed at later times, i.e., $z\leq 1$ (e.g., \cite{burke2012,montes2014,morishita2017,montes2017}), with the most dramatic evolution in the ICL fraction happening at $z\sim 0.5$ \citep{montes2017}. We estimated the ICL fraction in this sample using an accurate technique free of a priori assumptions called CICLE (CHEFs Intracluster Light Estimator, \cite{jimenez-teja2016}). CICLE is based on the use of Chebyshev-Fourier functions (CHEFs, \cite{jimenez-teja2012}) to model the surface light distribution of the galaxies and curvature maps to disentangle the ICL from the light of the BCG. CICLE studies the ICL two-dimensionally, without the need to simplify its surface distribution to a profile and, thus, not assuming any kind of symmetric distribution. It does not assume any previous hypothesis either, apart from the fact that the radial profiles of the BCG and the ICL must be different, i.e., analogous to having different ``slopes", which in practice is translated in surface analysis as having different curvatures. Notice that this is a minimum condition, since if it is violated no method would be able to disentangle the two light distributions. CICLE has been successfully tested with mock data and applied to real data from cluster Abell 2744 \citep{jimenez-teja2016}. \\

The test sample used in this work consists of eleven massive clusters with well defined dynamical states and similar observational characteristics, all with HST (Hubble Space Telescope) observations available. Most of the clusters are part of the Cluster Lensing and Supernovae Survey with Hubble (CLASH, \cite{postman}) and the Frontier Fields program (FF, \cite{lotz2017}). Both programs provided data of exceptional quality and depth, ideal for using CICLE. In this work we present the analysis of the ICL fraction for these eleven massive clusters in three different HST bands. We divided the sample into two groups, merging and relaxed clusters, aiming to study their {\it ICL fraction colors} (defined as the difference between two measurements of the ICL fraction of a cluster made at different wavelengths) and to unveil the possible progenitors of this ICL. The paper is organized as follows: Sect. \ref{sample_selection} describes the clusters in our sample and the criteria used to choose them. The observational characteristics of the data used and the preprocessing carried out is explained in Sect.  \ref{data}, while the CICLE algorithm is outlined in Sect. \ref{cicle}. In Sect. \ref{previous_results} we describe previous results regarding the ICL of the clusters in our sample, to compare them, to the extent possible, with the results that we obtain with CICLE in Sect. \ref{cicle_results}. Finally, we discuss our results in Sect. \ref{discussion} and draw the conclusions in Sect. \ref{conclusions}. Throughout the paper we assume a standard $\Lambda$CDM cosmology with $H_0$=70 km s$^{-1}$, $\Omega_m$=0.3, and $\Omega_{\Lambda}$=0.7. \\

\section{Sample selection} \label{sample_selection}

The data used for this work comes mainly from CLASH\footnote{http://www.stsci.edu/$\sim$postman/CLASH/Home.html} \citep{postman} and the FF program\footnote{http://www.stsci.edu/hst/campaigns/frontier-fields/FF-Data} \citep{lotz2017}, which includes both relaxed and merging massive clusters observed by the HST. \\

The CLASH program was mainly devoted to study the dark matter distribution in galaxy clusters using both strong and weak lensing and to search for Type Ia supernovae out to redshift $z\sim 2$ in 25 massive clusters especially selected. Twenty of these 25 systems were initially chosen to be likely relaxed, according to their symmetric surface brightness distribution in X-ray. The remaining five clusters were selected for being well known high magnification lenses. The cluster sample is distributed between $0.15<z<0.9$, and have masses between $5<M_{\mathrm{vir}}<30\,\times 10^{14}\,M_{\odot}$ \citep{postman,umetsu2014,merten2015}. All of them have X-ray temperatures $T_X>5$ keV. Each cluster was observed with both the ACS (Advanced Camera for Surveys) and the WFC3 (Wide Field Camera 3) in 16 passbands covering the NUV, optical, and NIR wavelengths \citep{postman}. \\

The FF (Frontier Fields) program (PI: Lotz) has observed with unprecedented depth six massive clusters using both ACS and WFC3 too. These systems were also chosen for being well known high magnification lenses, with the aim of studying not only the dark matter distribution in their cores but also analyzing the distant galaxies in the background, improving our knowledge of the Universe in the epoch of reionization. The six clusters, with redshifts ranging from $z\sim 0.3$ to $z\sim 0.55$ and masses spanning the interval $\sim 10<M_{\mathrm{vir}}<30\times 10^{14} M_{\odot}$, have been observed in seven different optical and NIR bands using 840 Hubble orbits \citep{lotz2017}.\\

Our sample is composed by six CLASH clusters ({\it Abell 383, Abell 611, MS2137-2353, MACS1115.9+0129, RXJ2129.7+0005} and {\it Abell 209}), four FF clusters ({\it Abell 2744, MACSJ0416.1-2403, MACSJ0717.5+3745}, and {\it MACSJ1149.5+2223}), and the western subcluster of the {\it Bullet} system. Although  {\it MACSJ0416.1-2403}, {\it MACSJ0717.5+3745}, and {\it MACSJ1149.5+2223} also belong to the CLASH sample, we used the FF images to analyze them, given that the CLASH data are included in them. Two more clusters were initially considered, the eastern subcluster in {\it Bullet} and the CLASH cluster {\it MACS1931.8-2635}, but due to the pollution from a nearby, bright star in the first case and the lack of enough spectra in the HST field-of-view in the second, these two systems were not included in the final study. The criteria to select these systems were: a) having similar masses, b) having high quality HST data available, c) having enough spectroscopic information available on the galaxies in the images, for the cluster membership determination, and d) having a well defined dynamical state, as determined by several indicators, if possible. The goal is to study consistently the ICL fraction with respect to the dynamical stage of the systems, using a homogeneous sample of objects under the same observational characteristics, with deep imaging data and with a significant number of precise galaxy redshifts. Splitting the sample according to the dynamical stage is thus crucial to disentangle which are the main mechanisms responsible for the ICL formation in each case. \\

The dynamical state of a cluster has been traditionally determined through visual inspection analyzing the morphology and the presence of substructure. Regular (relaxed) systems are, by definition, virialized, so they should be roughly circular, symmetric and without tidal features. They usually exhibit higher concentration indexes $c$ (here defined as the ratio between the light enclosed by a fixed inner aperture and the total light of the cluster) than unrelaxed systems \citep{cassano2010, donahue}. The existence of multiple BCGs is also an indicator of dynamical activity in a cluster related with its appearance. Furthermore, we also considered other parameters measured in X-rays, such as the symmetry  of the gas distribution. Dynamical interactions produce shocks or pressure waves that often break the symmetry of the gas distribution. Deviations from this symmetry are quantified through the {\it power ratio} and the {\it axial ratio}. The power ratio is a multipole decomposition of the X-ray surface brightness distribution that is sensitive to the presence of substructure, while the axial ratio is simply the ratio between the lengths of the minor and major axes of the X-ray distribution, thus providing an idea of its degree of elongation  (see \cite{cassano2010} and \cite{donahue} for a detailed description of these parameters). Also, the {\it centroid shift $w$}, defined as a statistical measurement of the projected offset between the X-ray peak and the centroid of the cluster measured within different circular apertures, serves to quantify the dynamical state of the system. It is expected that in relaxed clusters, the gravitational potential dominates the geometry of the system, making the hot X-ray emitting gas approximately align with the total mass distribution.\\

In the last decades, the presence of radio halos and radio relics has also been associated to merging clusters (e.g., \cite{cassano2010,cuciti2015,pandey-pommier, cassano2016}), especially in the case of massive systems,  through diffusive shock acceleration (e.g., \cite{ensslin1998}). Electrons in the intracluster gas are accelerated diffusively, using part of the energy dissipated during mergers in active clusters to energies where they would emit cluster scale ($\sim$Mpc-scale) synchrotron radiation. This emission appears in non-relaxed clusters in the form of {\it giant radio halos} or {\it giant radio relics}. Relatively passive systems with cool cores also exhibit diffuse synchrotron radio emission, but on smaller scales ($\sim$100 - 300 kpc) what is known as {\it radio mini-haloes}. In this case, the electron acceleration process is likely produced by AGN-driven turbulence in cool-core clusters (i.e., clusters with temperature profiles falling towards the center) causing this non-thermal emission \citep{brunetti} and is not related to the overall cluster dynamical state (e.g., \cite{bravi2016}). \\
 
Before describing the sample, it is important to keep in mind that it is much easier to tell if the cluster is merging than if it is relaxed. If there is no evidence of departures from relaxation we consider the clusters as relaxed. One should notice that the results of the analysis will not be dependent on the precise estimation of the relaxation level, but just that we compare clusters that have plenty evidence of merging with those that do not. We will now describe the properties of each one of the eleven clusters in our sample, in particular their dynamical states according to all the indicators previously described: \\

\begin{itemize}

\item {\it Abell 383} ({\it A383} hereafter, $z \sim 0.187$) is identified as a relaxed cluster according to the X-ray morphological parameters diagrams built by \cite{cassano2010} (see Fig. 1 in \cite{cassano2010} and Fig. 3 in \cite{donahue}). Its power ratios are very small and its X-ray distribution is highly circular, with an axis ratio of 
$\sim 0.97\pm0.01$ within a metric radius of 500 kpc \citep{donahue}. It has cool core with just a single point radio source of less than 5 kpc detected at the BCG \citep{giacintucci}. All evidence suggests that this is a regular system.

\item {\it Abell 611} ({\it A611} hereafter, $z \sim 0.288$) is also part of the CLASH sample displaying a circular and symmetric distribution in X-ray \citep{postman}. Although bright in X-ray, it is clearly relaxed, as the measurements of X-ray concentration, centroid shift, and power ratios made by \cite{donahue} indicate. No diffuse extended radio emission is detected, just central emission connected to the BCG \citep{venturi2008,pandey-pommier}.

\item {\it MS2137-2353} ({\it MS2137} hereafter, $z \sim 0.313$) appears to be a well relaxed cluster, as its X-ray morphological measurements suggest \citep{donahue}. We did not find any information on possible radio emission is available in the literature.

\item {\it MACS1115.9+0129} ({\it MACS1115} hereafter, $z \sim 0.352$) is a cool core cluster that is not as circular in X-ray as other clusters in the CLASH sample ($AR\sim0.85\pm 0.03$) but has a high X-ray concentration and small centroid shift and power ratios \citep{donahue}, all compatible with a relaxed state. A radio mini-halo was detected by \cite{kale2013,kale2015,pandey-pommier} and \cite{giacintucci}.

\item {\it RXJ2129.7+0005} ({\it RXJ2129} hereafter, $z \sim 0.234$) is also a clearly relaxed cluster according to its X-ray morphological properties, although its axis ratio is not as high as that from other systems in the CLASH "relaxed" sample ($AR\sim0.87\pm 0.01$) \citep{donahue}. It has a cool core. Its BCG hosts a strong radio source that is surrounded by a mini-halo \citep{kale2015,pandey-pommier,giacintucci}.

\item {\it Abell 209} ({\it A209} hereafter, $z \sim 0.206$) is a rich, X-ray luminous cluster \citep{mercurio2003a,mercurio2003b} that was originally selected as a relaxed cluster in the CLASH survey due to its symmetric X-ray distribution, although several indications of marginal departures from relaxation were also pointed out \citep{postman}. It is a non-cool core cluster, showing substructure in the galaxy velocity distribution and a marked luminosity segregation, strongly suggesting a merging state \citep{mercurio2003a,mercurio2003b}. Despite having small centroid shift and power ratios, its X-ray concentration is compatible with that of a merging cluster \citep{donahue}. The presence of a giant radio halo associated to its BCG \citep{venturi2007,pandey-pommier,giacintucci} seems to confirm that {\it A209} is in fact a non-relaxed cluster and it is either undergoing a merging event or at the end of a massive merger phase \citep{venturi2007,pandey-pommier}.

  \item {\it Abell 2744} ({\it A2744} hereafter, $z \sim 0.307$) is the first cluster observed by the FF program. It is a richness 3 cluster with a significant enhancement of the blue galaxy population (a blue fraction $\sim 2.2 \pm 0.3$ higher than that found in the same core regions of nearby clusters, \cite{couch&sharples1987}), mainly composed of starburst and post-starburst galaxies \citep{couch1998}. Analyzing combined X-ray and spectroscopic data, \cite{owers} identified two major substructures in the velocity distribution corresponding to the remnants of two major subclusters in a post-core-passage phase of merging with a large line-of-sight component, along with an interloping minor merger, model that was later confirmed and refined by \cite{merten2011}, who concluded that it is in fact a quadruple merging system. This result confirmed the previous works addressing the unrelaxed dynamical state of {\it A2744}, by \cite{kempner&david2004} and \cite{boschin2006}. It hosts a giant radio halo firstly found by \cite{giovannini} and later confirmed by \cite{venturi2008,kale2013} and \cite{giacintucci}, as well as a single radio relic in the outskirts \citep{govoni2001a,govoni2001b,kale2015}. All the evidence point to a heavily disturbed merging system.

\item {\it MACSJ0416.1-2403} ({\it MACS0416} hereafter, $z \sim 0.$396) is the most elongated cluster in the CLASH sample. It is a high magnification gravitational lens with an Ultra Steep Spectrum Radio halo (USSR) associated. It is the most powerful halo ever observed  \citep{pandey-pommier}.  Its power ratios, significantly higher than the average of the CLASH relaxed sample, also suggest a non-virialized state.  With low X-ray concentration and high centroid offset, it is clearly classified as a non-relaxed cluster according to the \cite{cassano2010} diagram \citep{donahue}. Moreover, there are several shifts between the peaks of lensing mass, the X-ray and radio emission, which, in conjunction with the presence of the USSR halo, point to an impressive four-cluster post-merging scenario similar to {\it A2744}, the Pandora cluster \citep{pandey-pommier}.

\item {\it MACSJ0717.5+3745} ({\it MACS0717} hereafter, $z \sim 0.$548)  is the CLASH system with the lowest X-ray concentration and highest centroid shift, showing also high dipole power ratios \citep{donahue}. It hosts a very powerful radio halo and a bright relic located in between the merging structures of the cluster \citep{vanweeren2009,pandey-pommier}.  An offset between the mass, X-ray, and radio distribution peaks is observed, which suggests that the system is an on-going merger in this case, since the steepness of the hosted radio halo is lower than that of {\it MACS0416} \citep{pandey-pommier}. 

\item {\it MACSJ1149..5+2223} ({\it MACS1149} hereafter, $z \sim 0.$544) is the MACS cluster with the highest velocity dispersion ($\sim 1800$ km/s). According to X-ray morphological diagrams by \cite{cassano2010} and the measurements by \cite{donahue}, this system is clearly classified as merging. The X-ray analysis performed by \cite{ogrean2016} confirms that this is a merging system with several substructures, displaying a line-of sight component, and with no evidence of a compact cool core. However, the lack of temperature substructures or surface brightness features, which would be expected in a such a complex merger, suggests that {\it MACS1149} is an old merger.  This scenario is supported by \cite{bonafede2012}, who reported evidence for a  giant, very steep radio halo,  as well as a double relic system, which could point to post violent merger phase \citep{pandey-pommier}. In addition, the dynamical analysis by \cite{golovich2016} confirms that this system seems to be composed by two different mergers involving three subclusters.

\item {\it 1E 0657-558} ({\it Bullet} hereafter, $z \sim 0.$296) has been thoroughly studied in the literature since its discovery in the nineties (e.g. \cite{barrena2002} and references hereafter). It is a textbook case for merging clusters and its nickname comes from the prominent Mach cone observed in X-rays, originated by a merger between two clusters very close to the plane of the sky \citep{markevitch2002, markevitch2004}. It is an ongoing merger where an infalling subcluster is observed just after its first core passage \citep{springel2007}. The presence of a radio halo was first noticed by \cite{liang2000}, and several other authors have confirmed this detection at different levels of significance (e.g., \cite{shimwell2014}).

\end{itemize}

\section{Data} \label{data}

We analyzed the  HST ACS images obtained in the CLASH and FF programs. The formidable depth and quality of the HST data is fundamental to study the ICL, given its low surface brightness of typically $\mu_V\geq26.5$ mag arcsec$^2$ (e.g. \cite{montes2014}). Deep ground-based data could also be used to detect and measure the ICL, although the fact that the absolute level of background contribution is larger makes the use of HST data preferable to decrease the uncertainty introduced by this component. Due to the reduced field-of-view of the ACS/WFC instrument (202 x 202 arcsec$^2$), some clusters had effectively imaged only its central area. CLASH data are available in eight optical ACS filters: F435W, F475W, F555W, F606W, F625W, F775W, F814W, and F850LP, whereas the FF images are observed in just three of them: F435W, F606W, and F814W. We decided to analyze the three common bands not only to compare the ICL fractions between merging and relaxed clusters at the same wavelength, but also to study if any statistical trends in ICL fraction colors could be identified. Detailed analyses of stellar populations in the ICL is beyond the scope of this paper and will be covered in a future paper, and that is why we do not attempt to quantify specific spectral features using other IR filters here. We have thus used the broadband F435W and F814W filters, and the F606W whenever it was possible. For one of the systems in our sample, {\it MS2137}, the data in the F606W band lacked the superb quality as compared to the rest of the observations and to maintain a similar quality level for the sample we decided to use the F625W filter instead. In any case, to test for possible biases due to this choice, for some clusters, the ICL fractions for these two intermediate wavelength filters have been estimated, as a proxy to show the difference in the ICL fraction between them. Although the FF data are deeper than those from CLASH, the excellent quality of the CLASH data, with a 5$\sigma$ limiting AB magnitude ranging from 27.2 (F435W) to 27.7 mag (F814W), guarantees that the ICL can be safely measured with these observations and compared to that obtained from the FF images.\\

For both CLASH and FF we used the combined, drizzled HST mosaics that have been created for these projects. These mosaics have been produced in a two-step process, where the first step in all cases begins with the individual raw exposures that are processed with the calibration pipeline {\it CALACS}\footnote{http://www.stsci.edu/hst/acs/performance/calacs\_cte/calacs\_cte.html}  at STScI, which includes corrections for bias, dark current, flatfield, non-linearity, charge transfer efficiency losses, and electronic gain and photometric calibration. For CLASH, these exposures were then subsequently aligned, corrected for geometric distortion, cosmic-ray rejected, and combined using the {\it MosaicDrizzle} pipeline \citep{koekemoer2002,koekemoer2011}, to produce mosaics with a pixel scale of $0.065^{\prime\prime}$per pixel. The FF mosaics were similarly processed, following the same techniques, with the pixel scale being $0.060^{\prime\prime}$ per pixel, and using the {\it Drizzlepac} software tools that were developed by \cite{gonzaga}. These fully-calibrated high-level products can be retrieved from the Mikulski Archive for Space Telescopes\footnote{https://archive.stsci.edu/prepds/frontier/}. Despite the fact that the Bullet cluster does not belong to the CLASH nor the FF samples, the CLASH team reduced and combined HST/ACS observations of this cluster too, putting them at the collaboration's disposal. For the western subcluster of the Bullet system we did not find observations in the F435W filter and this is the reason why we will not provide the ICL fraction measurement in this band. Even though the images in the three filters were available for the eastern subcluster, the contamination from a nearby, very bright star made it impossible to obtain reliable results, so that subcluster was excluded from the analysis. \\

As we need to measure the total luminosity of the cluster to estimate the ICL fraction, cluster galaxy members must be identified. Given the deep, high quality images of our clusters, we require spectroscopic information to identify the cluster members accurately and thus derive precise ICL fractions. We prefer just using spectroscopic redshifts to avoid the larger uncertainty and the interlopers that cluster membership algorithms based on photometric redshifts entail. Thus, the clusters in our sample were chosen for having not only reliable estimates of their  dynamical state, but also enough spectroscopic information. Table \ref{info_spec} shows the number of reliable spectroscopic redshifts publicly available for each cluster in our sample, as well as the source of these data. For those redshifts provided by NED\footnote{https://ned.ipac.caltech.edu/} (NASA/IPAC Extragalactic Database) we have rejected those with photometric or poor quality. Table \ref{info_spec} also shows the number of galaxy members confirmed by our two-step cluster membership algorithm (see Sect. \ref{cicle} and \cite{jimenez-teja2016} for further information on the classification procedure). Initially our sample also included the CLASH cluster {\it MACS1931.8-2635}, but after applying the cluster membership criteria we discovered that too few galaxies with spectroscopic redshift laid on the field of view of the CLASH observations and, therefore, it was discarded.\\

In order to know how the use of spectroscopic redshifts affects our results, we will examine the limiting magnitudes of the spectroscopic surveys used in this work. Given that for each cluster we have different sources contributing to the final spectroscopic sample (see Table  \ref{info_spec}), with the deepest data completing the previous spectroscopic surveys, we have determined that the worst case scenario is presented by the cluster {MACS0717} (excluding the {\it Bullet} cluster). This cluster not only has the shallowest spectroscopic sample ($r<21.2$ in the restframe, \cite{ebeling2014}) but also is the one with the highest redshift ($z\sim 0.548$). Calculating the distance modulus for this redshift, $DM=42.50$, we determine that our spectroscopic sample is complete up to an absolute magnitude of $M_r=21.2-42.50=-21.3$ mag for cluster {\it MACS0717}. Analyzing the luminosity functions calculated by \cite{connor2017} using photometric redshifts and assuming a completeness and purity of 100\% for the cluster galaxy members so derived, we estimate that our total luminosity might be underestimated by a $\sim 19.7\%$, for our worst case cluster. For the best system in our sample, cluster {\it A383} which has the lowest redshift and spectra from VLT/VIMOS, a similar analysis yields an underestimation of its total luminosity of $\sim 0.2\%$. We will analyze in Sect. \ref{cicle_results} the impact of this possible underestimation in the final ICL fractions.\\

\begin{deluxetable}{lcccl}
\tablecaption{Spectroscopic redshift information available for each cluster, number of galaxy members after applying the cluster membership algorithm, and bibliographic sources.}
\tablewidth{0pt}
\tablehead{
\colhead{Cluster} & \colhead{{\it z}} & \colhead{\# Spectra} & \colhead{\# Members} & \colhead{Source}}
\startdata
A383 & 0.187 & 1420 & 254 & 1, 2 \\
A611 & 0.288 & 1202 & 158 & 1, 3, 4 \\
MS2137 & 0.313 & 1874 & 408 & 1, 2 \\
MACS1115 & 0.352 & 1681 & 487 & 1, 2, 4 \\
RXJ2129 & 0.234 & 1654 & 184 & 3 \\
A209 & 0.206 & 1037 & 528 & 1, 2 \\
A2744 & 0.307 & 1518 & 348 & 1,7 \\
MACS0416 & 0.396 & 4386 & 643 & 6 \\
MACS0717 & 0.548 & 1267 & 581 & 1, 5 \\
MACS1149 & 0.544 & 617 & 311 & 1, 5 \\
Bullet (eastern) & 0.296 & 112 & 64 & 1 \\
\enddata
\tablecomments{(1) NED, (2) VLT/VIMOS, (3) Hectospec, (4) SDSS/BOSS, (5) \cite{ebeling2014}, (6) \cite{balestra2016}, and (7) \cite{owers}.}
\end{deluxetable}\label{info_spec}

\section{CICLE} \label{cicle}

In \cite{jimenez-teja2016} a new algorithm to study the ICL and estimate the ICL fraction was developed, called CICLE (CHEFs IntraCluster Light Estimator). The motivation was the need of finding a reliable and efficient algorithm able to disentangle the ICL from the light in cluster galaxies without assuming any a priori hypothesis. Traditional methods assume certain characteristics of the ICL, such as its surface brightness, density, distance to the brightest galaxies in the cluster, a symmetric morphology, or a certain radial profile. Instead, CICLE uses the CHEFs (Chebyshev-Fourier bases, \cite{jimenez-teja2012}) to model the light surface distribution of all the galaxies in the image, to obtain ICL and  background maps.\\

The CHEFs are mathematically orthonormal bases optimized to fit the two-dimensional light distribution of the galaxies. They are built in polar coordinates using Chebyshev rational functions to model the radial coordinate and Fourier modes to expand the angular component. The Chebyshev rational functions inherit the excellent properties of the Chebyshev polynomials, which have proven to provide optimal interpolations for smooth functions \citep{mason,boyd}. In addition, these approximations are readily computed, since the Chebyshev basis is very compact and we just need a few components to fit a profile with high precision \citep{jimenez-teja2012}. The Fourier modes, perfect to interpolate periodic functions, make it possible to fit any galaxy morphology, without requiring rotational symmetry. Thus, CICLE creates a two-dimensional CHEF model for every galaxy in a cluster image, including the BCG, to later remove it. Stars are just masked out.\\

Although fitting a regular galaxy is straightforward for the CHEFs, the case of the BCG is more complex, since it is difficult to know where the halo of the BCG ends and the ICL starts. For this reason, the BCG requires a differentiated treatment \citep{jimenez-teja2016}. After removing the CHEF models for all galaxies, CICLE re-inserts the CHEF model of the BCG. In this way, this central area is completely restored and we obtain an image composed just by the BCG, the ICL, and background. To estimate the limits of the light belonging to the BCG, a curvature map is calculated for the entire image. The curvature parameter is a characteristic of each pixel, and it represents the change in slope of a surface at a certain point in every direction. CICLE only assumes that the BCG and the ICL profiles have different slopes, otherwise it could be impossible to disentangle them. Under this assumption, the bi-dimensional limits of the BCG are defined by the points where the slope changes most, estimated through the curvature parameter. Once the outline of the BCG is computed, a new model for the BCG is built within the area delimited by its boundary. After removing this model, an image containing just ICL and background is obtained. We refer the reader to \cite{jimenez-teja2016} for a more detailed description of CICLE. \\

In \cite{jimenez-teja2016} the background level was estimated using images of  nearby fields, obtained approximately at the same epoch under the same observational characteristics. Since this is not possible for all the clusters in our current sample and in many cases there are no blank areas in the images to measure the background, we decided to use an homogeneous (albeit not so precise) approach, to be able to draw a consistent comparison between the final ICL fractions. We used the software SExtractor \citep{bertin} to estimate the background map for each cluster in our sample, using exactly the same background parameters. First, the background map is iteratively estimated  through  a $\kappa\sigma$-clipping algorithm in each mesh of a grid that covers the whole image. The size of each mesh is defined by the parameter BACK\_SIZE and it is one of the most influencing parameters in the final background estimation. As a general rule, BACK\_SIZE must be higher than the average size of the objects in the image, otherwise some flux from these objects could be absorbed in the background. It cannot be too large either, because small scale variations of the background would be erased. However, as we were trying to disentangle the background from the diffuse, extended light in the intracluster medium, small fluctuations are not as important as avoiding the contamination of light from the cluster galaxies and the ICL, so we set BACK\_SIZE = 512 \citep{holwerda}.\\

The grid of values yielded by this algorithm were later smoothed applying a median filter of size BACK\_FILTERSIZE. Again, as we were more interested in the average trend of the background rather than in small features, we chose a large filter of 5x5 pixels to smooth out any possible contamination from galaxies, stars and ICL. Then, this filtered grid was fit applying a bicubic-spline interpolation, which was later refined by recalculating the background locally around the objects in the image. For each object, the background was estimated in an annulus centered on it, with thickness set to BACKPHOTO\_SIZE = 24, which is the typical value for this parameter \citep{holwerda}. \\

With this approach we intended to obtain a background map for each cluster, estimated consistently to allow for a fair comparison of the final results. As can be noticed, our selection of values for the background related parameters was very different from the usual SExtractor configuration and some of them were intentionally high  with the aim of insuring, to the extent possible, that the background maps were smooth enough and did not contain any light from the galaxies or the ICL. \\

Finally, in order to measure the ICL fraction once we have a background-free ICL map, we created an image of the cluster removing the CHEF models of the foreground and background galaxies. As described in \cite{jimenez-teja2012}, the cluster membership is determined in a two-step process, the PEAK+GAP algorithm \citep{owers}, using the spectroscopic data available for each system. This composite method first identifies the peak of the cluster in the redshift space and selects a redshift window wide enough to contain the whole distribution of velocities assigned to that peak. Implicitly, the size of this window is proportional to the velocity dispersion of the clusters: merging clusters, with a more scattered velocity distribution, will need a wider window, compared to relaxed systems. This crude selection of cluster member candidates is obviously prone to contamination by interlopers. So, we further refine it using the shifting gapper method  \citep{fadda1996,girardi1996,boschin2006,owers}, which uses velocity and spatial information on the candidates simultaneously. The shifting gapper method distributes spatially the candidates according to their cluster-centric distance in radial bins. The mean velocity of the candidates within each bin is calculated and those candidates with velocities that are too far from the others are rejected. As unrelaxed clusters are more likely to have a broader spatial distribution, this procedure naturally allows candidates at larger distances to be identified as cluster members for these systems. These two steps are, thus, essential to guarantee that our cluster membership algorithm implicitly takes into account the dynamical state of the systems and does not bias the measurement of their total  luminosity, while minimizing contamination by interlopers at the same time. We refer the reader to \cite{jimenez-teja2012} for further information on the cluster membership selection algorithm.\\

\section{ICL in CLASH and FF clusters: previous results} \label{previous_results}

The properties of the ICL in CLASH and FF clusters have been already extensively studied by several authors using different techniques. We will briefly describe their results for different subsamples of the CLASH  \citep{presotto2014,demaio2015,burke2015,demaio2017} and FF clusters \citep{krick07,montes2014,morishita2017,montes2017}. The work by \cite{presotto2014}  was focused on the CLASH cluster {\it MACSJ1206.2-0847}, $z\sim 0.44$, which does not belong to our sample. They used deep multiband Subaru data to study the ICL properties. Light from cluster galaxies and fore- and background objects was modeled using traditional analytical profiles, such as single or double S\'ersic functions, masking out the galaxies with poor fits. To disentangle the BCG from the ICL they fit a composite de Vaucouleurs plus S\'ersic model, yielding a final ICL fraction of $4.3\pm 0.2\%$ at $R_{500}$ for the Rc band ($\lambda_0=6550$\AA). They compared this result with that obtained using a surface brightness threshold to identify the ICL, concluding that this method yields very different ICL fractions, depending on the value of this threshold, and systematically higher than that from the fitting technique. Assuming a  surface brightness level of $\mu_{R_c}=29.87$ mag/arcsec$^2$ (equivalent to $\mu_V=27.5$ mag/arcsec$^2$ at $z=0$) they got an ICL fraction of $4.7\pm0.4\%$.\\

\cite{demaio2015} studied the ICL in the IR for four CLASH clusters with $0.44\leq z\leq 0.57$, with {\it MACS1149} being the only one that we have in common in our sample. The ICL radial profile is measured from an ICL map obtained by masking out the light from the galaxies using either the SExtractor segmentation maps or by eye. Three of the clusters showed a significant radial gradient towards bluer color at larger cluster radii ({\it MACS1149} among them), interpreted as a gradient in metallicity assuming a fixed age for the ICL. The color of the ICL in the forth cluster was found to have a flat distribution. This study was later continued by \cite{demaio2017}, analyzing a larger sample composed by 20 clusters drawn from the CLASH set plus seven groups from the HST program \#12575, aiming to study the ICL colors and progenitors as a function of the halo mass. They obtained similar radial color gradients to those in \cite{demaio2015} and did not find statistical difference between clusters and groups. They did find a higher BCG+ICL mass fraction (assuming a fixed mass-to-light ratio) for groups than for clusters, as well as a more efficient ICL formation mechanism for low-mass halos within a radius of 100 kpc.  In both papers they concluded that the ICL formation is primarily driven by tidal stripping of the outskirts of massive galaxies ($M_{\star}>10^{10.4}M_{\odot}$). No ICL fractions were reported.\\

A subsample of 23 clusters from CLASH was analyzed by \cite{burke2015} using the technique of thresholding the surface brightness of the ICL ($\mu_B=25$ mag/arcsec$^2$) and masking out the stars and non-member galaxies with circles of radius proportional to their areas. They were able to estimate the ICL fraction in 13 out of the 23 clusters, reporting values between $\sim 2-23\%$. They find that their ICL fractions  strongly correlate with redshift, independently of the dynamical state of the clusters, growing by a factor of $\sim 4-5$ in $0.18\leq z\leq 0.90$ versus the $\sim 1.4$ growth factor found for the BCG from the accretion of its companions in the same redshift range. They conclude that the evolution of the ICL is mainly driven by minor mergers at low redshifts $z\sim 1$ as opposite to the BCG, which is primarily evolving at higher redshift.\\

\cite{montes2014} processed the images of the first FF cluster observed, {\it A2744}, using the rest-frame colors $g-r$ and $i-J$. They measured the ICL using three different estimates, the rest-frame surface brightness in the $J$ band the ICL ($\mu_J$), the logarithm of the stellar mass density ($\log(\rho)$), and the radial distance to the most massive galaxies of the cluster ($R$).  The corresponding thresholds established for each parameter were $24<\mu_J<25$ mag/arcsec$^2$, $\log(\rho)<1.2$, and $R>50$ kpc, yielding ICL fractions of 5.1\%, 4.0\%, and 10.5\%, respectively, within a radius of 400 kpc. They also used these two colors to study the properties of the stellar populations in the ICL, finding clear negative radial gradients for both age and metallicity towards the outskirts of the cluster. Their results suggest that the ICL in {\it A2744} is mainly formed by the disruption of infalling satellite galaxies with similar mass ($M_{\star}\sim 3\times 10^{10}M_{\odot}$) and metallicity than the Milky Way, being on average $\sim$5 Gyr younger than the most massive galaxies of the system. This cluster had its ICL fraction previously calculated by \cite{krick07}, using ground-based data observed by the du Pont 2.5m telescope in Las Campanas Observatory in two filters, the Gunn-$r$ ($\lambda_0=6550$ \AA) and $V$  ($\lambda_0=5400$ \AA). \cite{krick07} defined the ICL in these two bands using the rest-frame surface brightness thresholds of $\mu_r=26.4$ and $\mu_V=26.1$ mag/arcsec$^2$, respectively, yielding corresponding ICL fractions of $11\pm 5$ and $14\pm 5$ within one-quarter of virial radius. They found the ICL distribution to be multipeaked, with a color significantly redder than the red cluster sequence.\\

The work by \cite{montes2014} with {\it A2744} was later expanded to the whole FF sample in \cite{montes2017}. The ICL stellar population properties are defined using a distance criteria, assuming that the ICL is the luminous component that extends beyond a radius of 50 kpc once the galaxies in the image are masked using the segmentation maps provided by SExtractor. Under this definition, they found that the metallicity of the ICL for these six clusters is subsolar on average, and that its mean stellar age is between 2 to 6 Gyr younger than the most massive galaxies in the systems. They confirmed the stripping of $M_{\star}>10^{10}M_{\odot}$ galaxies to be presumably the principal driver of the ICL formation, occurring at $z<1$. To measure the ICL fraction they followed an approach similar to that developed for the {\it A2744} cluster, setting a surface brightness threshold of $\mu_V=26$ mag/arcsec$^2$ which yields ICL fractions in the range of $\sim 1-4\%$ in the V-band for all the FF clusters. In order to include the effect of the ICL flux that lies (in projection) within the area dominated by the BCG (defined as $r<50$ kpc), \cite{montes2017} linearly interpolated their ICL surface brightness profiles. The new ICL fractions ranged  between $\sim 4.8$ and $13\%$ within the $R_{500}$ radius, with  a mean of $\sim 7\%$, with their most relaxed cluster presenting marginal  evidence of having a higher ICL fraction compared to the other systems in the sample. Contrarily to \cite{burke2015}, they did not find any correlation of the ICL fraction with redshift. \\

The properties of the ICL in the six FF clusters were also studied by \cite{morishita2017} but with a completely different method. The brightest galaxies in the images ($m_{F160W}<26$) were fitted using single S\'ersic profiles plus a constant sky component in fixed-size ``postage stamps'' of 300 x 300 pixels. The constant sky level within these boxes was identified with the local ICL. A global ICL map was built as the weighted mean of all the overlapping boxes. The BCG did not receive a differentiated treatment, as its light is disentangled from the ICL as any other luminous galaxy in the field. Analyzing the colors of these ICL maps out to $R\leq 300$ kpc, they observed a radial gradient towards blue at larger cluster radii, as in previous studies. They also calculated the stellar mass distribution of the ICL, finding that the ICL is primarily dominated by moderately old stellar populations between $\sim 1-3$ Gyr old that could have been stripped from quiescent cluster galaxies with $M_{\star}<10^{9.5}M_{\odot}$ plus a $\sim 5-10\%$ fraction of younger stars (A- and earlier-type, $\sim 1$ Gyr) at $R\leq150$ kpc presumably coming from recently star-forming/infalling galaxies. Although ICL light fractions are not computed, they reported ICL mass fractions calculated from SED fitting for all the six clusters, ranging between $\sim 7-23\%$ within a radius of 300 kpc and $\sim 4-19\%$ for $R\leq 500$ kpc. Again, no trend can be identified with redshift, as in \cite{montes2017}.\\

ICL or ICL+background maps of CLASH and FF clusters have also been obtained as by-products in other works with photometric or gravitational lensing purposes although neither the ICL properties nor the ICL fraction are studied in these works (e.g., \cite{merlin2016,livermore2017,molino2017,connor2017}). \\

\section{Results with CICLE} \label{cicle_results}

We estimated the ICL fraction of our sample of 11 massive clusters in three ACS/WFC broad bands (F435W, F606W, and F814W), whenever possible. In the case of {\it MS2137} the F606W observation from CLASH was not as good as the rest of the data, so we decided to process the F625W filter instead. We also estimated the ICL fraction in the F625W filter (as well as in the F606W band) for one of the merging clusters ({\it MACS0416}) and one of the relaxed systems ({\it A383}), just for comparison. In the case of the {\it Bullet} cluster, we only estimated the ICL fractions for the eastern subcluster, since the measurements for the western subcluster were polluted by the presence of a nearby, very bright star. Unfortunately, we did not have data for the eastern subcluster in the F435W filters, and that is the reason why we report the ICL fractions just in the F606W and F814W bands.\\

The images were first preprocessed to mask out the brightest stars, since they are not smooth enough to be processed by the CHEFs \citep{jimenez-teja2016}. We masked the areas associated to these stars in the SExtractor segmentation map for the F814W band, enhanced using a 10x10 pixel filter. The same mask is applied to the other two filters F435W and F606W, to ensure that the differences in the ICL fractions are physical and not induced by different masking. These masked pixels are excluded from the final measurement of the ICL fraction.\\

Then we run CICLE to obtain ICL+background maps (see Sect. \ref{cicle}).  The original images and the resulting maps for each cluster in the different filters are displayed in Appendix \ref{ICLmaps}: Figs. \ref{a383} to \ref{rxj2129} for the relaxed subsample, and Figs. \ref{a209} to \ref{bullet} for the unrelaxed systems. The typical value of the background found for our whole sample ranges from approximately (7 to 8)e-05 cps for the three main filters considered, completely consistent with the values calculated by \cite{morishita2017} using a different algorithm. These background values represent the $\sim 33\%$ (F435W), $\sim 9.0\%$ (F606W), and $\sim 8.5\%$ (F814W) of the ICL flux. Using the $r_{200}$ radii reported by \cite{boschin2006,maier2016,martinet2017} and \cite{morishita2017}, we calculated the apparent size of the clusters in our sample. We found no correlation between the background values measured in the three main filters and the apparent size of the clusters, guarantying that our background measurements are not biased.   \\

\begin{deluxetable}{lCCCCCCCC}
\tablecaption{Results yielded by CICLE for the 11 clusters in our sample: ICL fractions and errors for the different filters and the radii used to measure these ICL fractions. Systems belonging to the relaxed sample are marked with an asterisk.}
\tablecolumns{9}
\tablehead{\colhead{} & \multicolumn{2}{c}{F435W} & \multicolumn{2}{c}{F606W} & \multicolumn{2}{c}{F625W} & \multicolumn{2}{c}{F814W}\\ \hline
\colhead{Cluster} & \colhead{ICL fraction} & \colhead{Radius} & \colhead{ICL fraction} & \colhead{Radius} & \colhead{ICL fraction} & \colhead{Radius} & \colhead{ICL fraction} & \colhead{Radius}\\
\colhead{} & \colhead{[\%]} & \colhead{[kpc]} & \colhead{[\%]} & \colhead{[kpc]} & \colhead{[\%]} & \colhead{[kpc]} & \colhead{[\%]} & \colhead{[kpc]}}
\startdata
A383* & 11.16\pm0.77 & 63.6 & 8.25\pm2.18 & 104.0 & 10.06\pm2.84 & 94.7 & 6.18\pm5.33 & 108.2\\
A611* &  7.48\pm3.98 & 141.1 & 7.22\pm1.26 & 159.1 & & & 9.41\pm0.95 & 252.5\\
MS2137* & 9.48\pm0.71 & 135.4 &  & & 7.16\pm2.99 & 179.6 & 4.86\pm2.80 & 316.0\\
MACS1115* & 7.29\pm5.79 & 163.8 & 9.52\pm2.13 & 252.0 & & & 10.99\pm3.79 & 250.5\\
RXJ2129* & 2.95\pm3.74 & 63.9 & 10.26\pm0.31 & 227.6 & & &  7.95\pm7.53 & 176.2\\
A209 &  13.45\pm0.67 & 128.6 & 18.03\pm3.57 & 312.2 & & & 17.24\pm4.04 & 276.5\\
A2744 & 16.23\pm0.78 & 183.8 & 19.95\pm3.06 & 288.4 & & & 19.30\pm1.18 & 330.8 \\
MACS0416 & 15.12\pm0.22 & 336.8 & 22.78\pm0.19 & 328.3 & 19.90\pm0.51 & 310.6 & 11.29\pm1.60 & 332.8 \\
MACS0717 & 7.22\pm0.81 & 275.3 & 22.27\pm3.68 & 562.5 & & & 13.63\pm3.60 & 421.6 \\
MACS1149 &  11.90\pm1.34 & 172.5 & 20.52\pm2.24 & 336.2 & & & 18.39\pm5.91 & 626.3\\
Bullet (eastern) & & & 20.64\pm7.35 & 217.9 & & & 12.00\pm1.11 & 349.5 \\
\enddata
\end{deluxetable}\label{ICLresults}

We then computed the radial flux profiles of the ICL surface. As in \cite{jimenez-teja2016}, these radial profiles are obtained by averaging the flux inside the natural contours of the ICL in the core of the cluster and inside ellipses in the outskirts. These radial profiles show a negative slope reaching a minimum from which the ICL flux starts to increase due to the instrumental light from the borders of the images, or where the ICL submerges into the background. We thus measured the ICL fraction up to that radius, where the flux profile is minimum, and beyond which our estimation would be unreliable  due to spurious instrumental effects, as described in \cite{jimenez-teja2016}. Given the similar depths in the filters F606W and F814W for the CLASH data ($\sim 27.6$ and $\sim 27.7$ AB mag for a 5$\sigma$ point source within a $0.4^{\prime\prime}$ diameter aperture, respectively \citep{postman}), we can presume that difference in the ICL radii between these two filters is physical. However, the F435W depth is $\sim 27.2$ AB mag, reason why observational characteristics might be the cause of the different areas. As for the FF images, F435W and F606W depths are virtually the same $\sim 28.8$ AB mag while filter F814W is slightly deeper ($\sim 29.1$ AB mag) \citep{lotz2017,merlin2016}.\\

We summarize in Table \ref{ICLresults} the resulting ICL fractions and the corresponding radii of the measurements. The errors associated to the ICL fractions were estimated as the quadratic sum of the photometric error of the measured flux and the intrinsic error of the CICLE algorithm in the disentanglement of the BCG from the ICL. The former error is negligible in comparison to the latter in most of the cases, due to the high signal-to-noise of the HST images. However, as we described in Sect. \ref{data}, the use of spectroscopic redshifts for the cluster membership could cause a possible underestimation of $\sim 19.7\%$ in the total luminosity of the cluster, for the worst system in our sample (excluding the {\it Bullet} cluster). Propagating the errors and using the highest  ICL fraction found for this cluster ($22.27\%$, which yields the largest error associated to this problem), this would imply, if anything, a maximum error of $\sim 4.39\%$ to be added to the values listed in Table \ref{ICLresults}. This upper limit in the error induced by the limiting magnitude of the spectra confirms the advantage of the use of these data, whenever available, instead of photometric information for the cluster membership, since it is small compared with the potential contamination introduced by photometric redshifts in this identification. For the sake of comparison, the underestimation of the total luminosity for our best case cluster {\it A383} is approximately $\sim 0.2\%$, which is translated into a an additional ICL fraction error of $\sim 0.02\%$, which is completely negligible.\\

As was described in \cite{jimenez-teja2016}, the second source of error, the error of the CICLE algorithm, was estimated using mock images with the same characteristics of the real data: for each image, we created a simulated image of the same size and containing two exponential profiles with effective radii and surface brightness equal to those of the real BCG and ICL surfaces. We then polluted the mock images with ten realizations of noise with the signal-to-noise of the original observations and applied CICLE to them. The final error was obtained as the mean of the errors of the ten realizations. \\

\begin{figure}
\centering
\includegraphics[width=16cm]{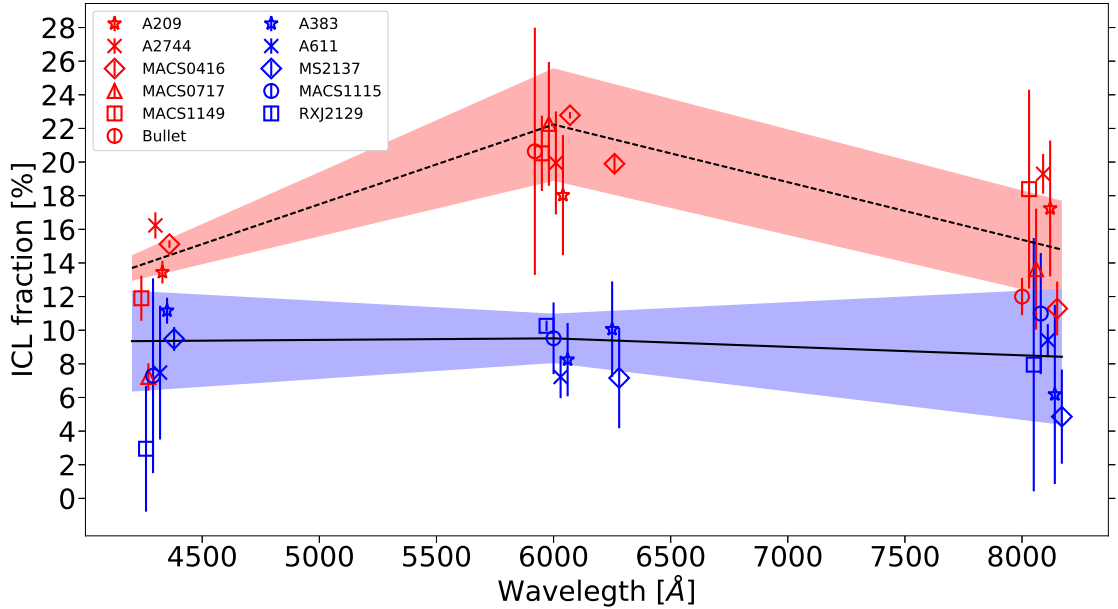}
\caption{ICL fractions yielded by CICLE for our sample of eleven clusters. Red markers represent merging clusters while blue markers are associated to relaxed systems. The black lines indicates the error weighted mean for each subsample (solid for relaxed clusters and dashed for merging systems), and the shaded areas represent the mean of the errors. For clarity, we have offset horizontally the points by 30 \AA~gaps. }\label{plotICL}
\end{figure}

The obtained ICL fractions are plotted in Fig. \ref{plotICL}, where the merging clusters are represented with red markers and the regular systems with blue symbols. For clarity, measurements for some clusters have been slightly offset horizontally by 30 \AA~gaps. We observe that the relaxed systems seem to have  a nearly constant gradient in the ICL fraction within the errorbars, while the disturbed clusters surprisingly show a clear increase in the F606W ICL fraction. Although on average the ICL fractions of the merging sample are higher than those of the regular clusters, we conclude from Fig \ref{plotICL} that the bluest and the reddest filters, F435W and F814W, cannot be use to discriminate between relaxed and non-relaxed systems. However, the ICL fraction in the F606W broadband quantitatively describes a significant difference between the dynamical states of the clusters.\\

Due to the disparate techniques applied, the different reference filters used, and the different apertures to estimate the ICL fraction, it is difficult to have a direct quantitative comparison of the ICL fractions estimated by us to those in the literature (see \cite{rudick2011} for a comparison of methods). Our ICL fractions are, in general,  comparable or higher than those reported previously. This is partially explained by the fact that CICLE includes in the estimations the ICL projected over the BCG-dominated area, and this is not the case for surface brightness- and radial distance-based methods. By design,  in these traditional techniques the ICL projected over the central regions is not added up to the final estimation of ICL flux since these pixels are excluded. This lost flux is of great importance, since the ICL is known to be more concentrated in the central area, and this may cause a significant underestimation of the ICL fraction \citep{willman2004, rudick2006}. Moreover, our ICL fractions are not measured homogeneously up to the same metric radius, but we restrict our calculations to the total area of  the aperture that is not contaminated by spurious instrumental light. In many cases our  radii are smaller than those used in the literature, thus yielding higher ICL fractions \citep{rudick2006}.  \\

We do not observe any trend in the ICL fraction with redshift, in contrast with \cite{burke2015} but in line with the works by \cite{krick07, montes2017} and \cite{morishita2017}. One should notice that, as we are working with ICL fractions and not ICL fluxes, the results shown are robust without the need for applying any correction for redshift or evolution.\\

\section{Discussion} \label{discussion}

A first attempt to link the dynamical state of the clusters with the properties of the ICL was made by \cite{feldmeier2002}, although without quantifying the amount of ICL or the ICL fraction. Similarly to us, \cite{krick07} obtained lower ICL fluxes for apparently more relaxed systems when they analyzed a sample of ten clusters in the redshift range $0.05\leq z\leq0.3$. Using the M3-M1 parameter (magnitude difference between the first- (M1) and the third- (M3) brightest galaxy members of the cluster) as an estimator of the dynamical age of the system, they found that it correlates with the ICL luminosities measured in the broadband  Gunn-r band (with a central wavelength $\lambda_0=6550$ \AA~ which can be considered comparable to our ACS F606W filter). They also observed similar relations for other dynamical indicators, as for instance, the M2-M1 magnitude difference (defined in a similar way as the M3-M1), or simply the presence of single, large elliptical galaxies (cD) in the cluster core. Moreover, the ICL fluxes found for clusters without cD galaxies are twice as high as  those measured in systems with cD galaxies on average. Interestingly, in spite of using a different sample, data with different observational characteristics, and a different technique, yielding lower ICL fractions than ours, \cite{krick07} observe the same trend. However, as the timescales governing the dynamics of the stars in the ICL are not consistent with settling down into the the center of the gravitational potential (or BCG) due to angular momentum or energy losses \citep{merritt1984}, \cite{krick07} conclude that either stars in the ICL have been formed early in groups which migrate to the center, or they have been directly stripped at the center of the cluster potential at later times, or even that the ICL is observationally indistinguishable from the BCG halo. They also find steeper ICL profiles for relaxed clusters, which would favor the hypothesis of an ICL evolution linked to the BCG formation: as groups are being merged with the BCG, they could bring their primordial  ICL stars with them and, at the same time, create even more ICL by ram-pressure of the gas or dynamical friction. However, in the absence of a central dominant galaxy,  groups and cluster galaxies would evolve slowly by tidal forces and dynamical friction, barely influencing the stars in the ICL and displaying shallower ICL profiles.  \\

Moreover, whereas the energy and angular momentum of the groups dissipate and they bring their ICL stars with them, any ICL formed by galaxy interactions would stay in the orbit where it was formed \citep{krick07}.  If this mechanism for producing ICL can be efficient at larger radii this would explain the ICL radial color gradient reported by several authors (e.g. \cite{montes2014,demaio2017}), in particular for the systems in the CLASH and FF samples. \cite{demaio2017} found that 75\% of the ICL luminosity in the CLASH clusters was consistent in color with the stars stripped from the outskirts of  cluster galaxies with $M_{\star}>10^{10.4}M_{\odot}$,  while \cite{morishita2017} observed 90-95\% of the ICL mass in the FF systems had colors that were compatible with the outer regions of quiescent cluster galaxies of $M_{\star}<10^{9.5}M_{\odot}$. Then, if the lower-metallicity/younger stars stripped from the outskirts of the luminous galaxies stay in their orbits, they will create the bluer trend towards larger cluster radii observed. This gradient would also be expected from the contribution to the ICL from the disruption of low-mass, low-metallicity dwarf galaxies, which are completely shredded at larger cluster radii compared to more massive, more metal-rich galaxies \citep{demaio2015,demaio2017}. \\

Numerical simulations predict a growth in the ICL fraction with decreasing redshift \citep{willman2004,rudick2006, rudick2011}. However, some authors found that the ICL fraction changes slightly over short timescales (as major mergers or collisions occurs). For instance,  \cite{willman2004} predict that the amount of ICL is directly linked to the infall of large groups already containing unbounded ICL stars, although they also find that this does not necessarily changes the ICL fraction. \cite{rudick2006} made a very detailed study of the effect of the dynamics on the ICL fraction in three different clusters, concluding that the ICL fraction growth is mainly driven by accretion events of massive galaxies and groups falling into the cluster center. For those systems that did not experienced major interactions, the ICL fraction evolved passively, rising slowly. However, it is interesting that, although the amount of ICL increases dramatically in mergers (potentially even doubling the its luminosity), they observed a decline in the ICL fraction at the beginning of the interaction. This is explained by their definition of ICL through a surface brightness threshold of $\mu_V=26.5$ mag/arcsec$^2$, which biases their measurements of the ICL fraction in the time preceding a major merger. As the galaxy groups start to infall, the luminosity appears more concentrated and thus temporarily increases the surface brightness of the cluster. Stars belonging to the ICL are therefore boosted and misclassified as part of the cluster galaxies during a short time. At later merging epochs, they did observe a rise in the ICL fraction associated to the merger event, in general agreement with our results. This conclusion was later corroborated by other authors with different numerical simulations (e.g. \cite{murante,contini}). \\

In general, \cite{rudick2011} proved that the different definitions of ICL fraction show consistent behaviors, in spite of yielding very different values for the ICL fraction . In particular, the ICL fraction was found to rise irregularly during the cluster evolution due to merging events, which can easily cause the ICL fraction in individual clusters to deviate from the global average trend with redshift.  This implies that the ICL fraction alone in a single filter cannot be a robust estimator of the dynamical stage of the clusters, since a merging cluster at higher redshift could have a similar ICL fraction than a relaxed system at lower redshift. However, our findings  raise the possibility of using the ICL fraction color instead of the ICL fraction to estimate the dynamical stage of clusters, at least in the case of massive systems.\\ 

\begin{figure}
\centering
\includegraphics[width=16cm]{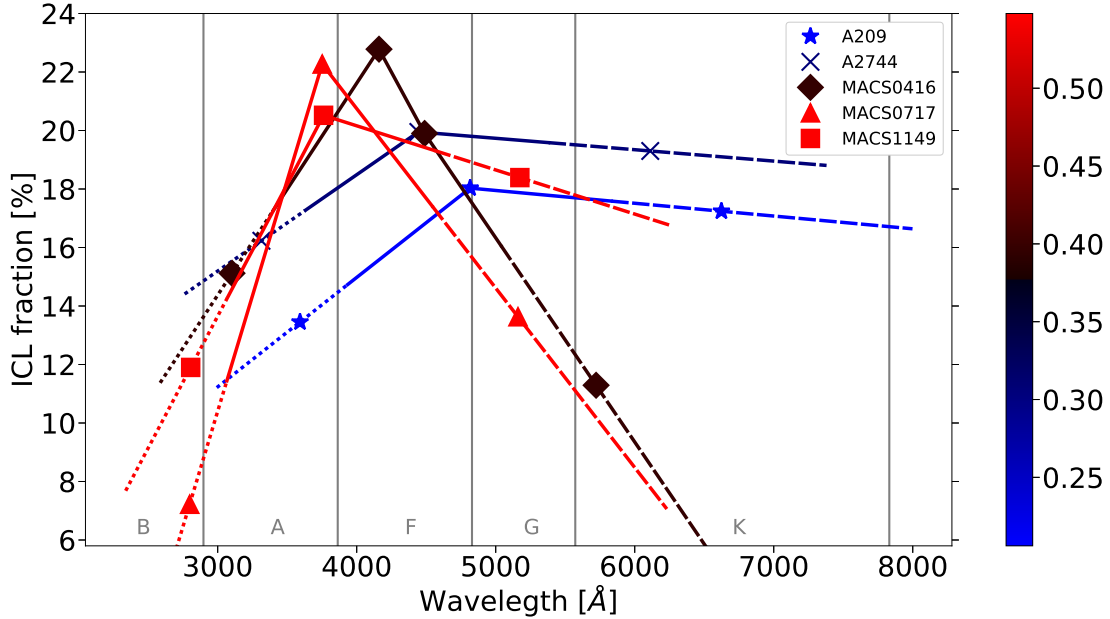}
\caption{ICL fractions yielded by CICLE for our subsample of merging clusters at rest-frame wavelength. Lines are color coded by redshift and different styles are used to represent the wavelength range covered by each one of the three filters: dotted for the F435W filter, solid for the F606W filter, and dashed for the F814W filter. Vertical gray lines separate the wavelength intervals were the emission peaks of the different stellar spectral types are included, as indicated at the bottom of each region.}\label{merging_results}
\end{figure}

For our subsample of relaxed systems, we observe that the distribution of the ICL fractions along the different wavelengths is nearly constant within the error bars except for the case of the cluster {\it RXJ2129} (see Fig. \ref{plotICL}). That means that the colors of the ICL are coincident on average with those of the stellar populations in the galaxies, considering the cluster as a whole. This is consistent with the idea that these systems have reached a virialized stage and the ICL stellar populations are just evolving passively and the ICL fraction is slowly fed by the stars stripped out from the  cluster member galaxies by dynamical friction. However, for the subsample of merging clusters the ICL fraction exhibits a strong increase in the F606W band, significant at $2.5\sigma$ with respect to the subsample of relaxed systems. Compared to the increase observed in the other two filters, this excess is especially significant with respect to the F435W filter, given the smaller errorbars. So, for clusters suffering major merger events, we see an excess of flux for the ICL in the F606W band compared to the cluster galaxies light, meaning that a significant fraction of bluer stars, presumably with lower metallicities, is being  stripped out violently from the outskirts of the infall galaxies. The presence of these stars \citep{goddard2017a,goddard2017b} would cause the ICL to be bluer than the overall light from galaxies, in comparison to relaxed clusters. An interesting question is why this blueing-merger effect is more pronounced systematically in the F606W filter rather than in the F435 band. In their detailed analysis of the properties of the ICL in the six FF clusters, \cite{morishita2017} found that the ICL was mainly composed (in mass) by moderately old stellar populations ($\sim 1-3$ Gyr) which would contribute more to the F606W filter than to the F435W. However, they also observed a non-negligible fraction of the ICL stellar mass that was likely associated to a bluer/younger population ($\sim 1$ Gyr). They estimated that approximately a $\sim 5-10\%$ of the ICL mass was compound of A- or earlier-type stars, probably stripped from star-forming galaxies during the cluster merging process. A-type stars have a lifetime of $\sim 1$ Gyr on average which, compared to an average crossing time ($\sim 1$ Gyr) would make possible to see the influence of these stars on the ICL fraction.  In Fig. \ref{merging_results} we can visualize the filters where A-type stars flux would contribute the most, according to the redshift of each cluster. The ICL fractions are now plotted at the rest-frame wavelengths, color coded by the redshift of the clusters, and the line style indicates the wavelength range covered by each filter: dotted lines for the F435W band, solid line for the F606W filter, and dashed line for the F814W band.  Given that A-type stars display temperatures from 7500 to 10000 K, their peak emission will range from $\sim 2900$ to 3900 \AA~. For the two highest redshift clusters in our sample {\it MACS1149} and {\it MACS0717}, at $z\sim 0.544$ and $z\sim 0.548$ respectively, this emission would be almost completely included in the F606W filter, with little contribution to the F435W flux. For the merging system {\it MACS0416} at $z\sim 0.396$, the young population flux contribution would be divided between the two filters. However, for the lowest redshift FF cluster in our sample, {\it A2744} ($z\sim 0.307$), A-type stars would be mostly observed in the F435W band, which could presumably explain why the gradient between the F606W and the F435W filter in this cluster is not as pronounced as for the rest of the merging subsample. Although the cluster {\it A209} does not belong to the FF sample and we do not have information on the possible ICL stellar populations, its low redshift  ($z\sim 0.206$) and ICL fraction colors coincident with those of {\it A2744} suggest a similar explanation based on the presence of younger stars.\\

We must also notice that, in spite of having ICL fractions consistent with those of our relaxed sample, the regular cluster {\it RXJ2129} displays an ICL fraction color distribution behavior similar to that of the merging sample, with a peak in the F606W. Even though the errors estimated for the F435W and F606W measurements are high we can  presume that the ongoing minor mergers pointed out by other authors \citep{kale2015, pandey-pommier, giacintucci} could be the origin of this fluctuation. This would be an interesting prediction for the relation between dynamical activity in clusters and the color (SED) gradient described in this work.  If this is confirmed, the color distribution of individual clusters ICL fraction could be used to estimate the mass ratio of mergers solely using optical data. \\

The relation between ICL fraction color gradient and cluster dynamics described in this work can be extremely useful not just to select clusters for further analysis of the merging process, but also to exclude merging clusters from scaling relations for mass proxies used in cosmology with purely optical data such as those incoming from the current and near future mega-surveys, such as DES\footnote{www.darkenergysurvey.org} and J-PAS\footnote{www.j-pas.org}.\\

\section{Conclusions} \label{conclusions}

We have analyzed the ICL in eleven systems with high quality imaging and enough spectroscopic information available, with the aim of characterizing their dynamical state through the ICL fraction. We have applied CICLE, a new algorithm described in \cite{jimenez-teja2016}, that is free of a priori assumptions on the properties both of ICL and galaxies. In that work, CICLE was found to estimate the ICL fraction with a maximum error of 10\% in the absence of noise for reasonable configurations of the ICL+BCG system. CICLE was thus proved to be robust and accurate, ideal to consistently study our sample of eleven HST-observed clusters.\\

The dynamical stage of the clusters was carefully determined compiling the different results available in the literature, gathering probes on X-ray morphology, dynamical analysis, and radio information. Five of the systems had strong indications to be relaxed systems, while the other six showed clear signs of dynamical activity. The resulting ICL fractions in the three optical broad band filters F435W, F606W, and F814W for the subsample of regular clusters were nearly constant within the error bars, ranging between $\sim 2-11\%$. For the six merging clusters, we report higher ICL fractions on average in the three filters, in the interval $\sim 7-23\%$. A different behavior is displayed observing the ICL fraction colors, with a significant peak in the ICL fraction measured at the intermediate band. Both the higher ICL fractions and the peak at the F606W band are consistent with previous results in the literature, although derived from simulations or analyses of the ICL colors. No obvious trend is identified in the ICL fraction with redshift.\\

Although a larger sample of galaxy clusters with clearly defined dynamical states and HST-like observational characteristics is necessary to improve our statistical significance, we have shown that the ICL fraction colors, measured robustly and consistently, can offer valuable information on the dynamical processes occurring in clusters.  Since bonafide relaxed systems are more difficult to be classified as such than merging systems, the addition of a truly relaxed system ICL fraction measurement would be extremely desirable to establish the range variation of this color gradient with respect to merger stage. In that case systems very old and relaxed such as fossil groups of galaxies with deep enough observations at comparable redshifts would be the best candidates, which are currently unavailable in the HST archive.   \\

 \acknowledgments
 
We thank the referee for constructive comments that helped to improve the original manuscript. Y. J-T. would like to thank Dr. Marc Postman for his support, help, and encouragement during her stay at the STScI, which made possible this work. We gratefully acknowledge the computational support of Dr. Fernando Roig. Y. J-T. also acknowledges finantial support from the Funda\c{c}\~ao Carlos Chagas Filho de Amparo \`a Pesquisa do Estado do Rio de Janeiro - FAPERJ (fellowship Nota 10, PDR-10) through grant E-26/202.835/2016, and  the Coordena\c{c}\~ao de Aperfei\c{c}oamento de Pessoal de N\'ivel Superior - CAPES (Science without Borders program, Young Talent Fellowship, BJT) through grant A062/2013.  R.A.D. acknowledges support from the Conselho Nacional de Desenvolvimento Cient\'ifico e Tecnol\'ogico - CNPq through BP grant  312307/2015-2,  and the Financiadora de Estudos e Projetos - FINEP grant  REF. 1217/13 - 01.13.0279.00. Both Y. J-T. and R.A.D. also acknowledge support from the Spanish National Research Council - CSIC (I-COOP+ 2016 program) through grant COOPB20263, and the Spanish Ministry of Economy, Industry, and Competitiveness - MINECO through grants AYA2013-48623-C2-1-P and AYA2016-81065-C2-1-P.

\appendix

\section{ICL+background maps} \label{ICLmaps}

\begin{figure*}[h]
\centering
\includegraphics[width=16cm]{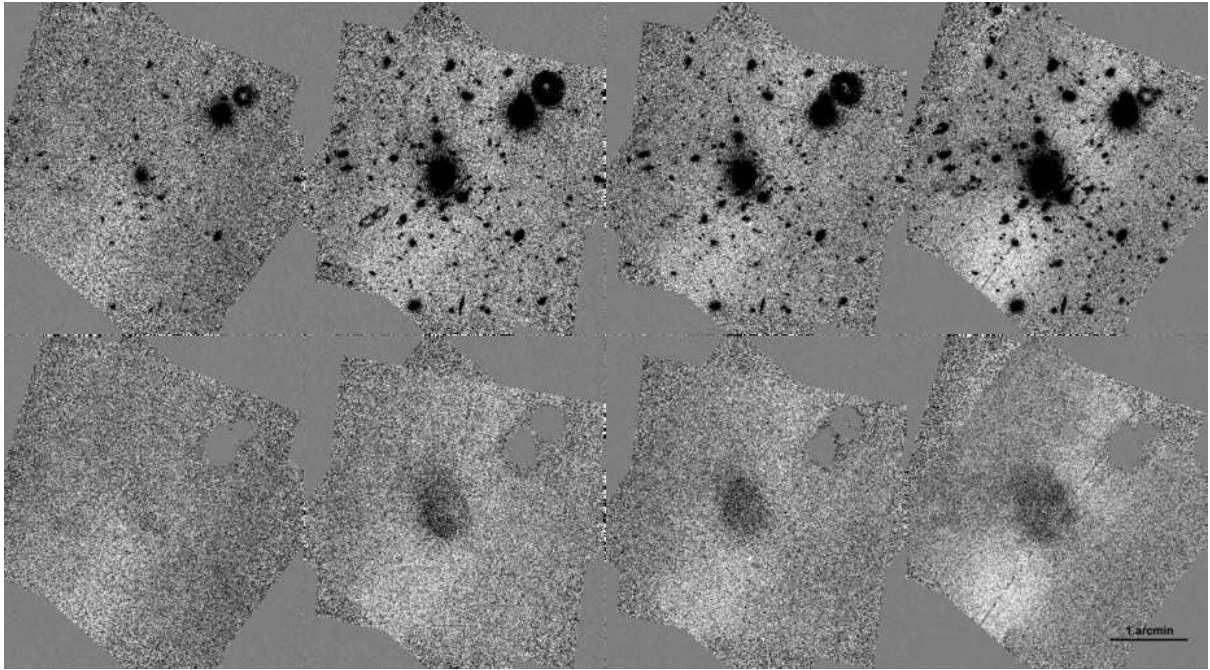}
\caption{Original images of the relaxed cluster {\it A383} (top) and ICL+background maps provided by CICLE (bottom) in the F435W, F606W, F625W, and f814W filters (from left to right). The scale of the original and ICL+background images is the same for each filter.}\label{a383}
\end{figure*}

\begin{figure*}[h]
\centering
\includegraphics[width=16cm]{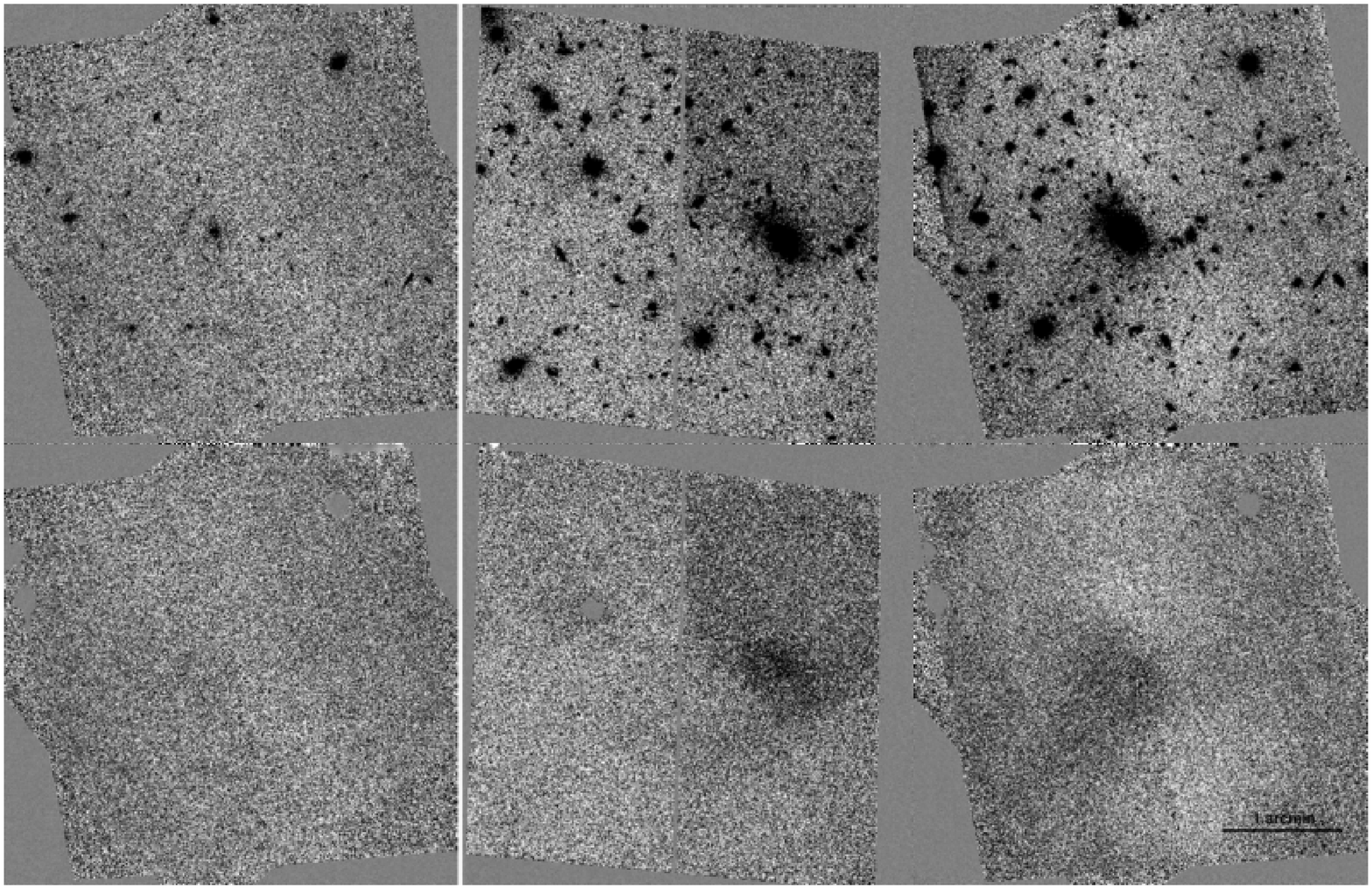}
\caption{Original images of the relaxed cluster {\it A611} (top) and ICL+background maps provided by CICLE (bottom) in the F435W, F606W, and f814W filters (from left to right). The scale of the original and ICL+background images is the same for each filter.}\label{a611}
\end{figure*}

\begin{figure*}[h]
\centering
\includegraphics[width=16cm]{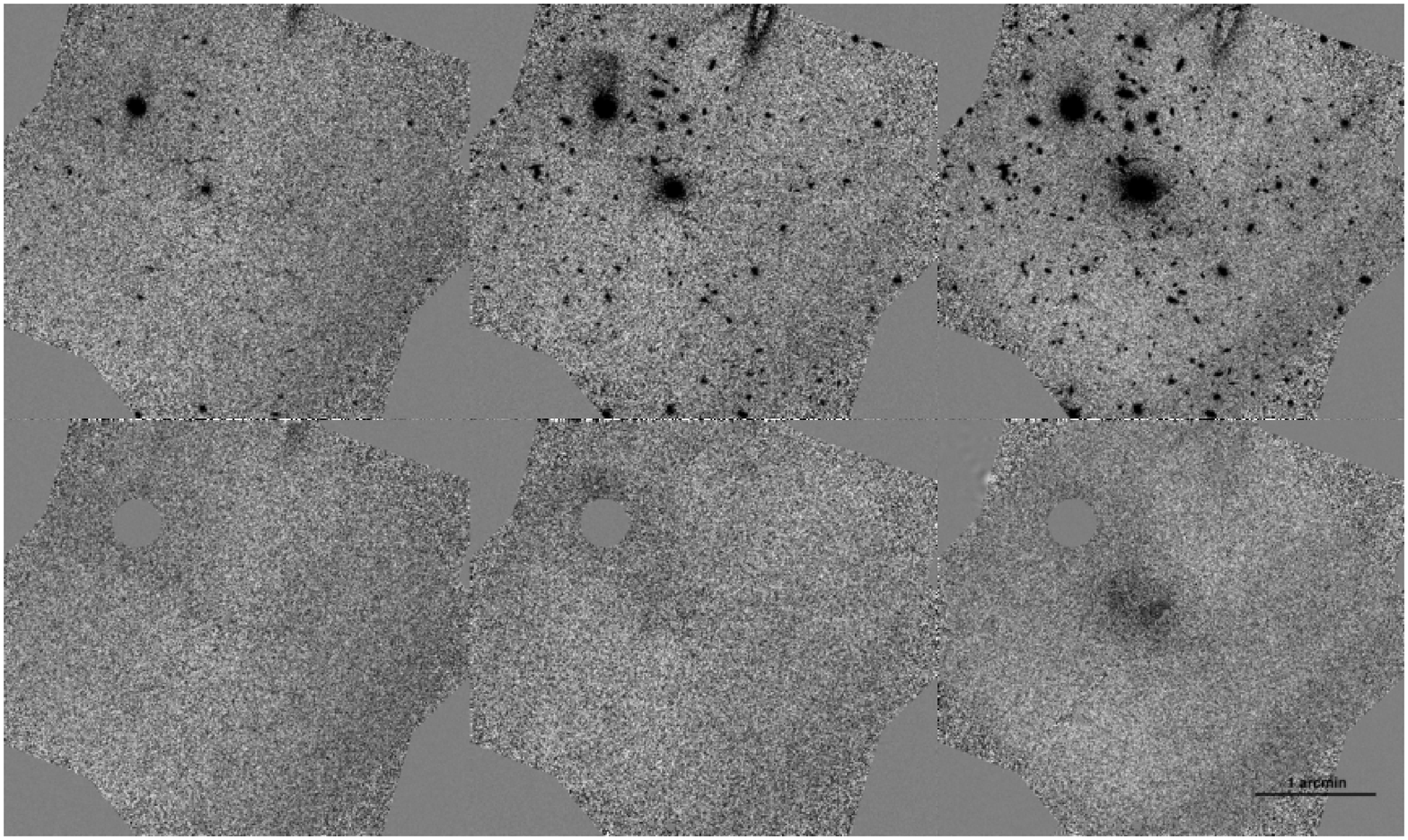}
\caption{Original images of the relaxed cluster {\it MS2137} (top) and ICL+background maps provided by CICLE (bottom) in the F435W, F625W, and f814W filters (from left to right). The scale of the original and ICL+background images is the same for each filter.}\label{ms2137}
\end{figure*}

\begin{figure*}[h]
\centering
\includegraphics[width=16cm]{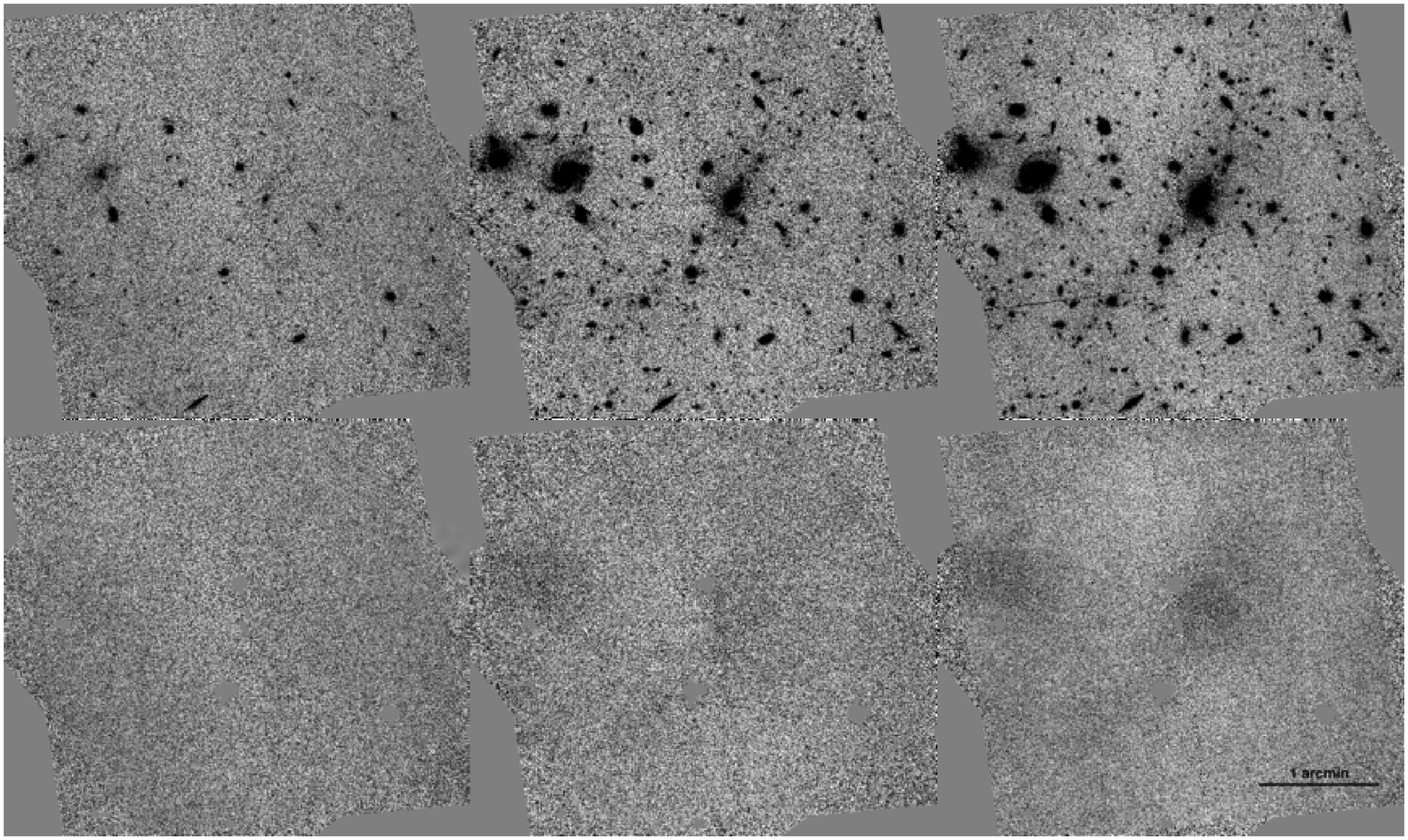}
\caption{Original images of the relaxed cluster {\it MACS1115} (top) and ICL+background maps provided by CICLE (bottom) in the F435W, F606W, and f814W filters (from left to right). The scale of the original and ICL+background images is the same for each filter.}\label{macs1115}
\end{figure*}

\begin{figure*}[h]
\centering
\includegraphics[width=16cm]{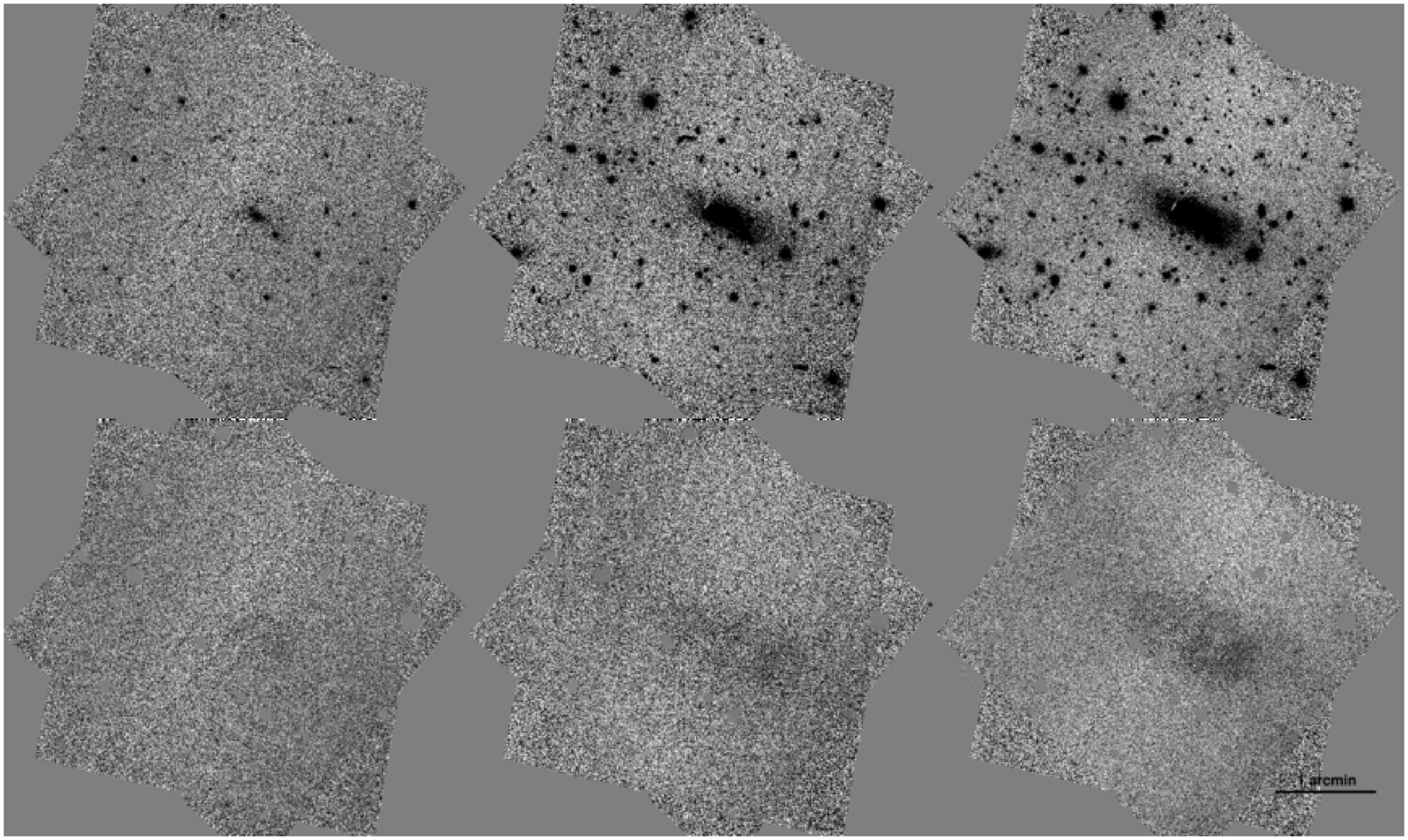}
\caption{Original images of the relaxed cluster {\it RXJ2129} (top) and ICL+background maps provided by CICLE (bottom) in the F435W, F606W, and f814W filters (from left to right). The scale of the original and ICL+background images is the same for each filter.}\label{rxj2129}
\end{figure*}

\begin{figure*}[h]
\centering
\includegraphics[width=16cm]{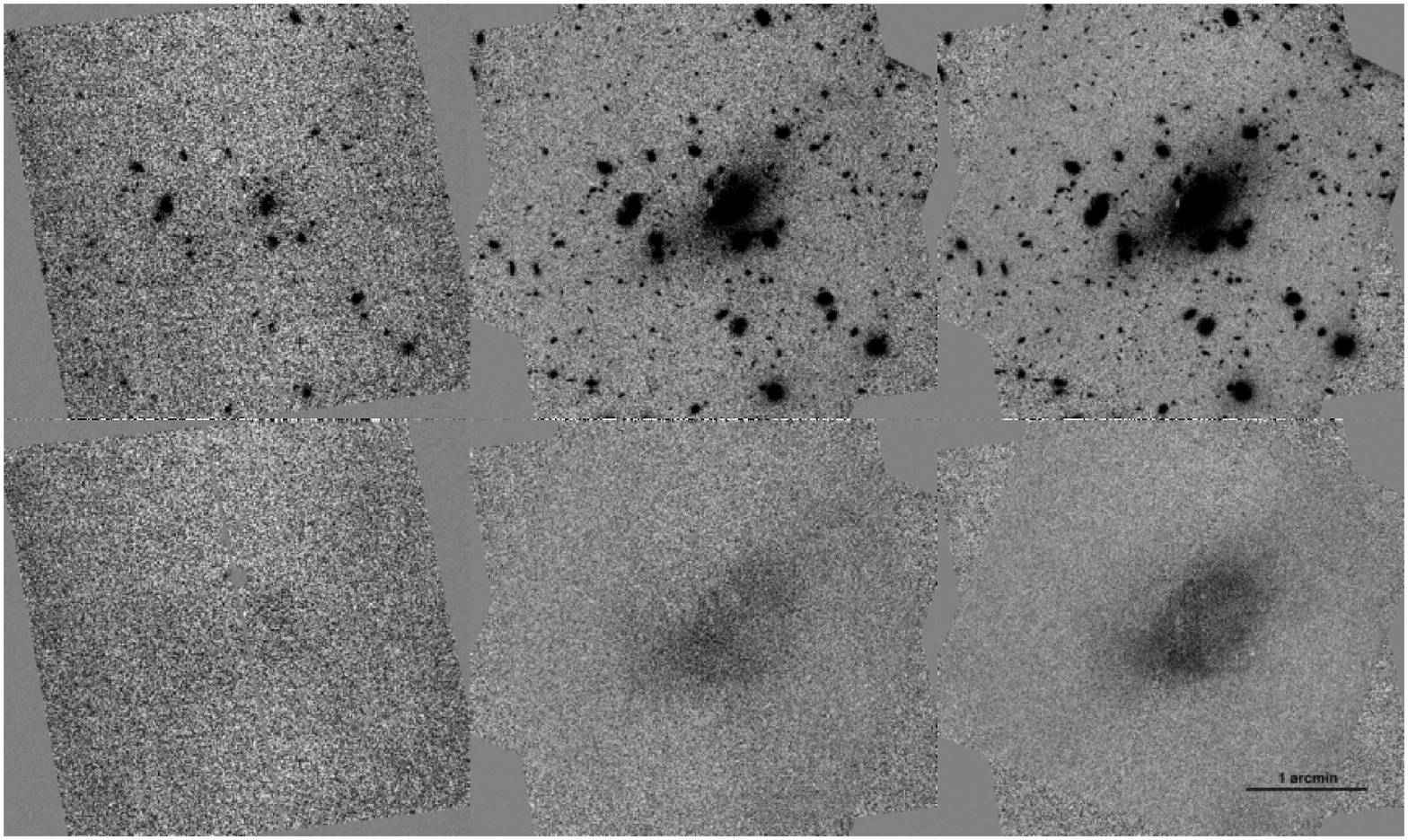}
\caption{Original images of the unrelaxed cluster {\it A209} (top) and ICL+background maps provided by CICLE (bottom) in the F435W, F606W, and f814W filters (from left to right). The scale of the original and ICL+background images is the same for each filter.}\label{a209}
\end{figure*}

\begin{figure*}[h]
\centering
\includegraphics[width=16cm]{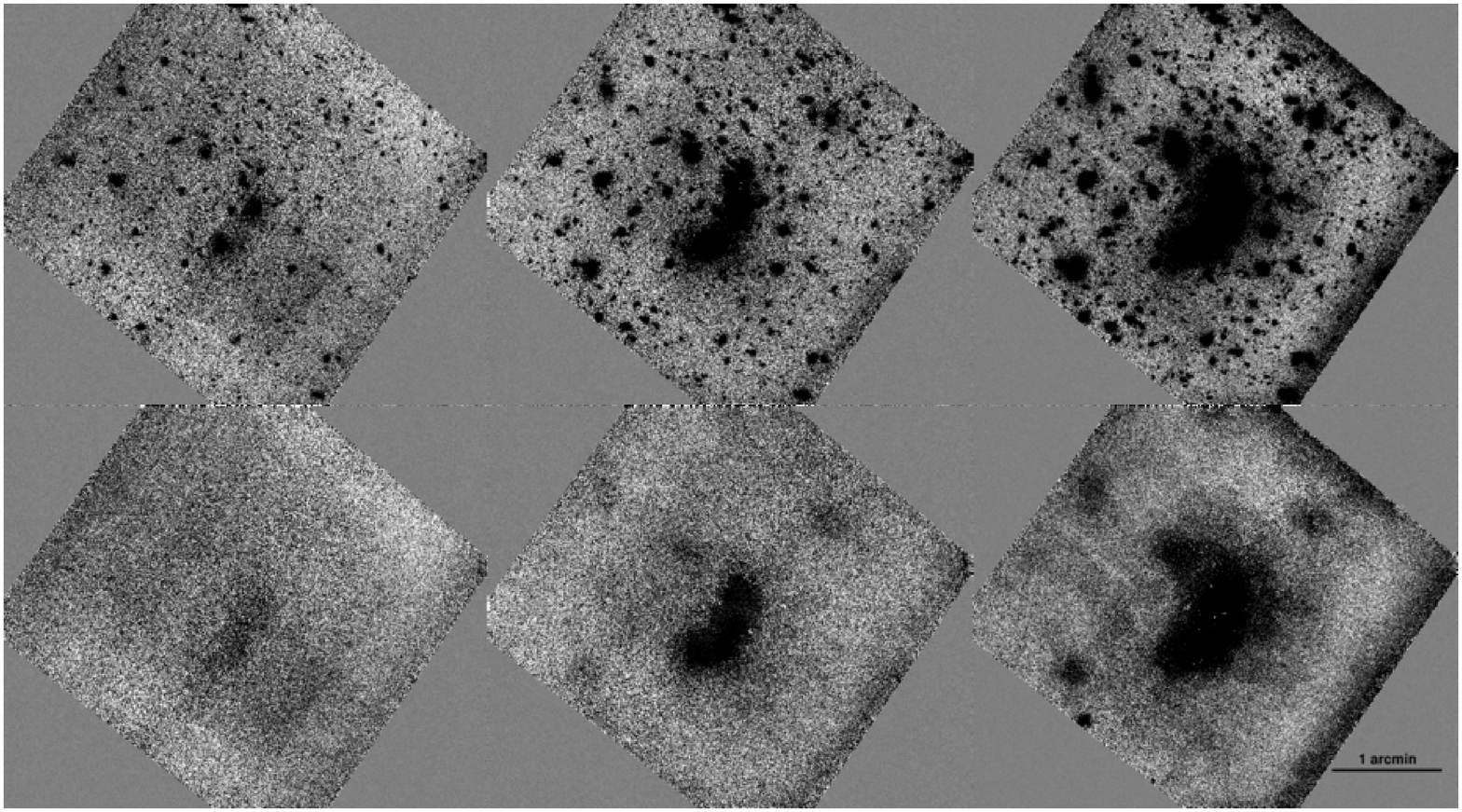}
\caption{Original images of the unrelaxed cluster {\it A2744} (top) and ICL+background maps provided by CICLE (bottom) in the F435W, F606W, and f814W filters (from left to right). The scale of the original and ICL+background images is the same for each filter.}\label{a2744}
\end{figure*}

\begin{figure*}[h]
\centering
\includegraphics[width=16cm]{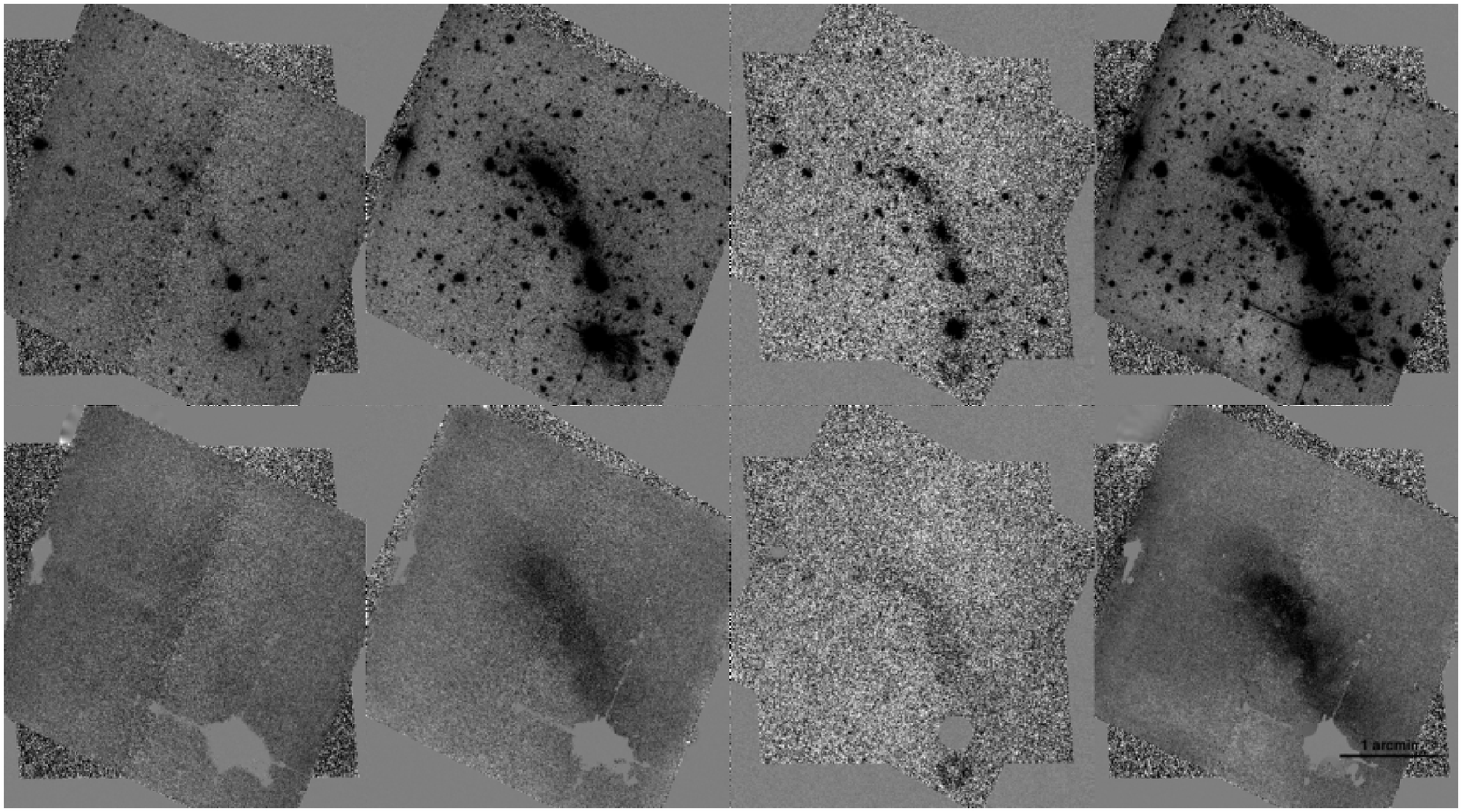}
\caption{Original images of the unrelaxed cluster {\it MACS0416} (top) and ICL+background maps provided by CICLE (bottom) in the F435W, F606W, F625W, and f814W filters (from left to right). The scale of the original and ICL+background images is the same for each filter.}\label{macs0416}
\end{figure*}

\begin{figure*}[h]
\centering
\includegraphics[width=16cm]{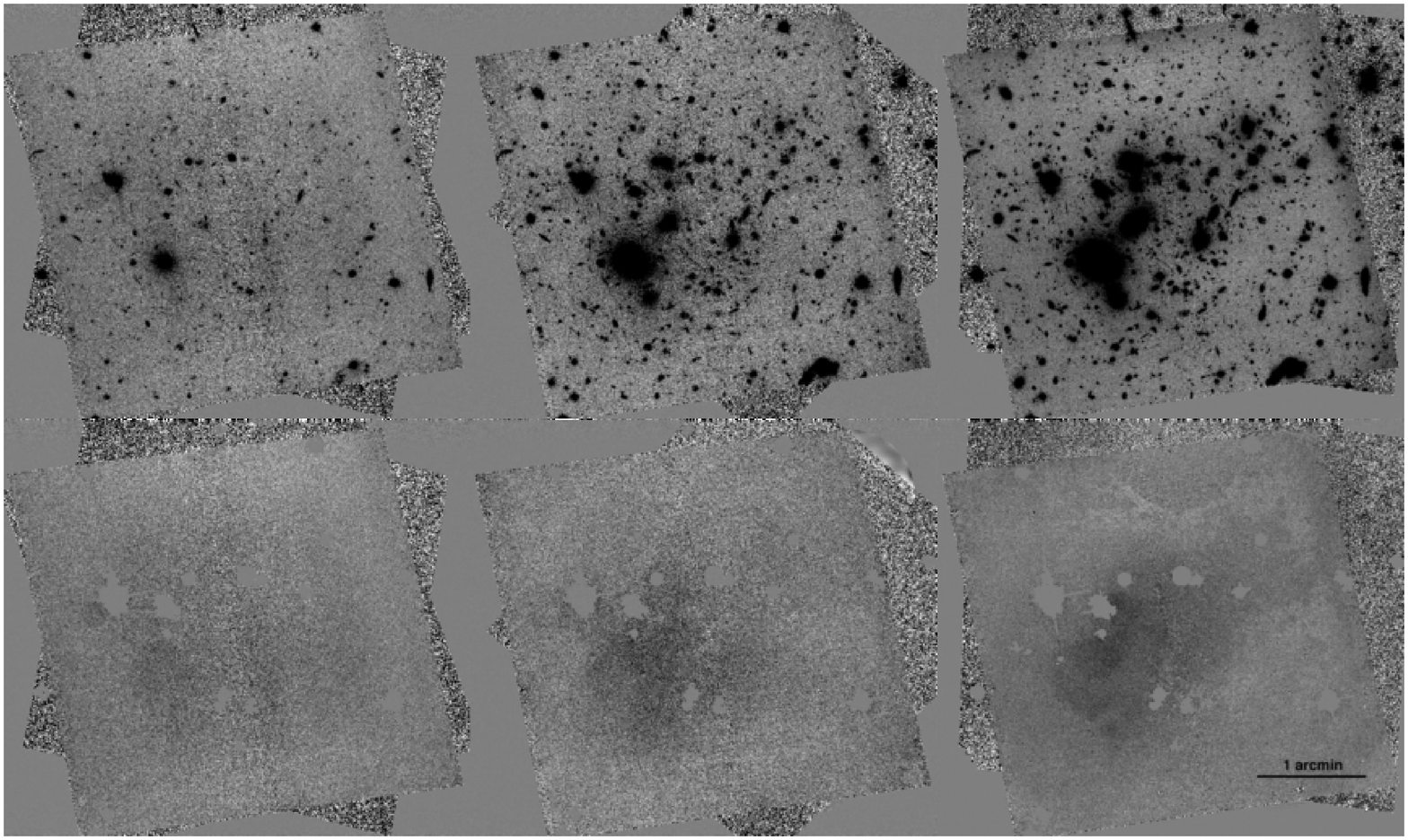}
\caption{Original images of the unrelaxed cluster {\it MACS0717} (top) and ICL+background maps provided by CICLE (bottom) in the F435W, F606W, and f814W filters (from left to right). The scale of the original and ICL+background images is the same for each filter.}\label{macs0717}
\end{figure*}

\begin{figure*}[h]
\centering
\includegraphics[width=16cm]{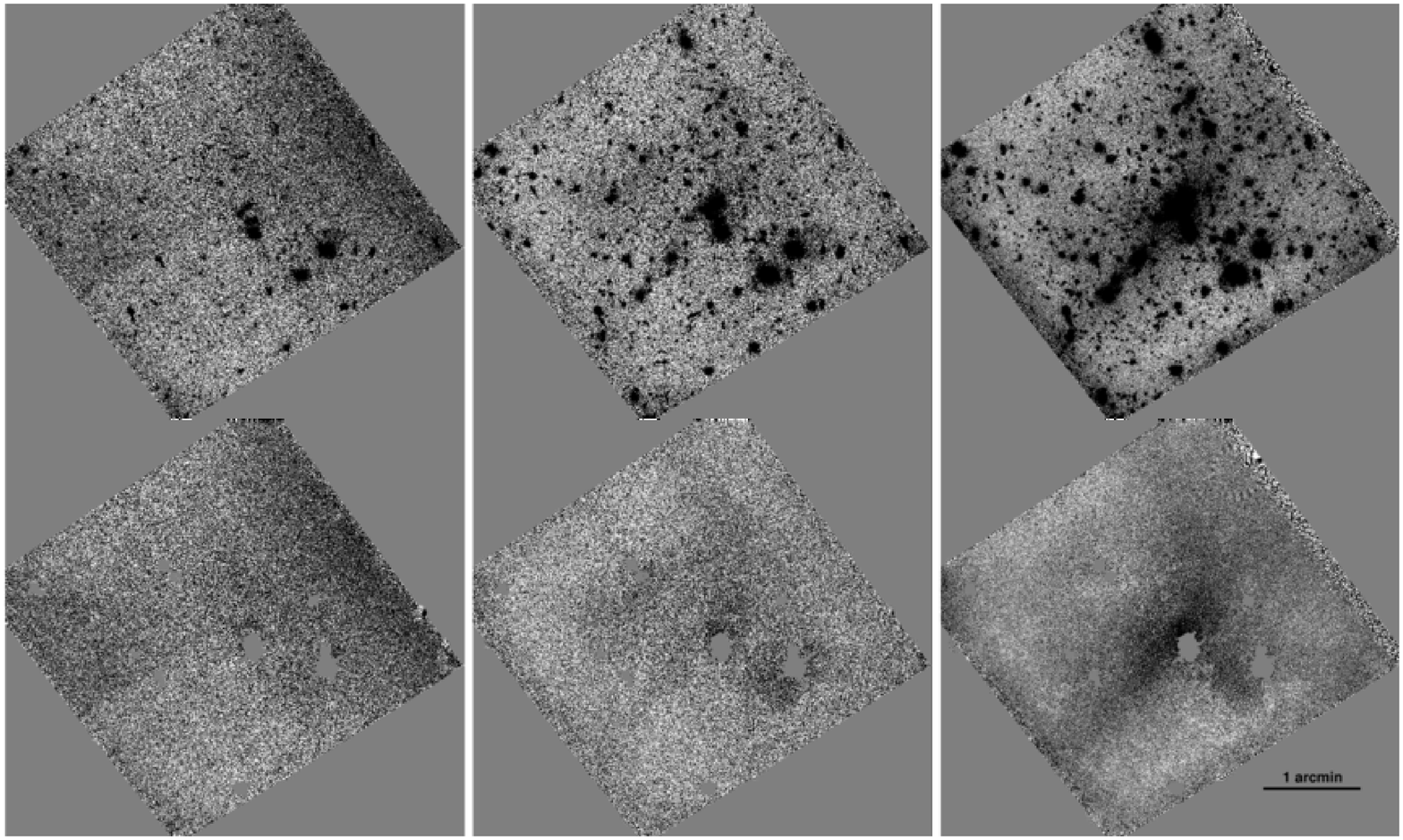}
\caption{Original images of the unrelaxed cluster {\it MACS1149} (top) and ICL+background maps provided by CICLE (bottom) in the F435W, F606W, and f814W filters (from left to right). The scale of the original and ICL+background images is the same for each filter.}\label{macs1149}
\end{figure*}

\begin{figure*}[h]
\centering
\includegraphics[width=16cm]{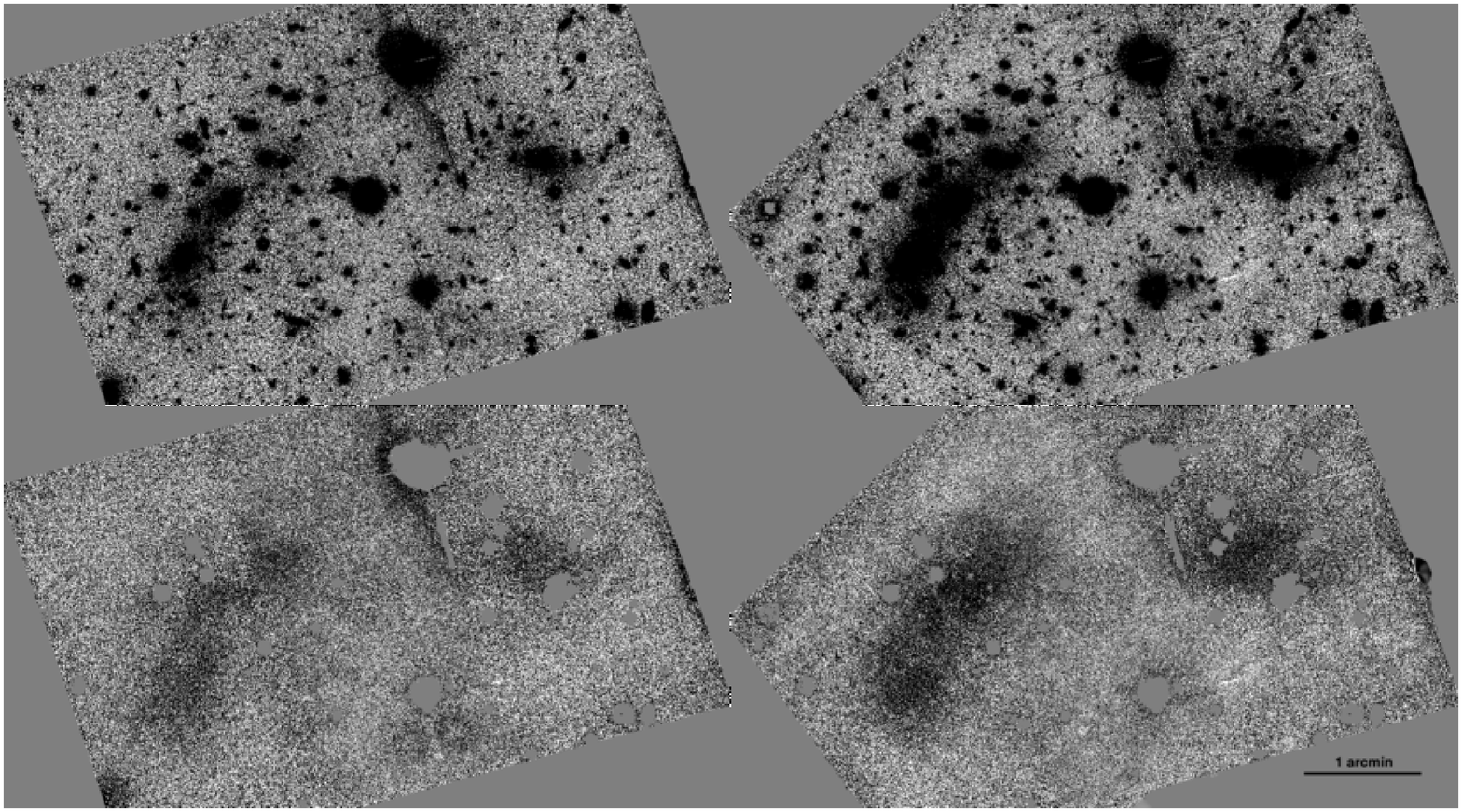}
\caption{Original images of the unrelaxed {\it Bullet} cluster (top) and ICL+background maps provided by CICLE (bottom) in the F606W and f814W filters (from left to right). The scale of the original and ICL+background images is the same for each filter.}\label{bullet}
\end{figure*}



\end{document}